\documentclass[letterpaper,fleqn,usenatbib]{mnras}

\usepackage{enumerate}
\usepackage{graphicx}	
\usepackage{amsmath}	

\def\eg{{\it e.g.}}
\def\etal{{\it et al.}}
\def\etc{{\it etc.}}
\def\ie{{\it i.e.}}
\def\Msun{M$_\odot$}
\def\funit{km~s$^{-1}$~kpc$^{-1}$}
\def\Ab{A$_{100}$}
\def\Cb{C$_{100}$}

\title[Thickening of Bars]{Three Mechanisms for Bar Thickening}

\author[Sellwood \& Gerhard]
          {J. A. Sellwood$^{1}$\thanks{E-mail:sellwood@as.arizona.edu}
and
{Ortwin Gerhard$^2$\thanks{E-mail:gerhard@mpe.mpg.de}}
\\
$^1$Steward Observatory, University of Arizona, 933 N Cherry Ave,
Tucson AZ 85722, USA \\ $^2$Max-Planck-Institut f\"ur Extraterrestrische Physik, Giessenbachstrasse, D-85748 Garching, Germany}

\pubyear{2020}

\begin{document}
\label{firstpage}
\pagerange{\pageref{firstpage}--\pageref{lastpage}}
\maketitle

\begin{abstract}
We present simulations of bar-unstable stellar discs in which the bars
thicken into box/peanut shapes.  Detailed analysis of the evolution of
each model revealed three different mechanisms for thickening the
bars.  The first mechanism is the well-known buckling instability, the
second is the vertical excitation of bar orbits by passage through the
2:1 vertical resonance, and the third is a gradually increasing
fraction of bar orbits trapped into this resonance.  Since bars in
many galaxies may have formed and thickened long ago, we have examined
the models for fossil evidence in the velocity distribution of stars
in the bar, finding a diagnostic to discriminate between a bar that
had buckled from the other two mechanisms.
\end{abstract}



\section{Introduction}
\label{sec.intro}
Many disc galaxies possess bars \citep[\eg][]{Er18}.  A bar is usually
the strongest non-axisymmetric feature in the light distribution
\citep[\eg][]{DG16} and is often, but not always, centered on the
photometric and kinematic centre of the
galaxy.\footnote{Counter-examples are in generally low-mass galaxies,
  such as the LMC \citep[\eg][]{dVF72} and NGC~1313
  \citep[\eg][]{CA89}.}  The material that makes up a bar manifests
considerable non-circular streaming motion \citep{Ko83, We01, Ag15,
  Ho15}, indicating a strongly non-axisymmetric gravitational field.
Many galaxies that are viewed edge-on manifest a boxy or peanut shaped
bulge \citep{Sh87, Lu00, ED17}, which is interpreted \citep{CS81, KM95,
  BA05} to be a bar that has become thicker than the surrounding disc.

The Milky Way possesses a strong bar seen in oblique projection from
the Sun's location \citep{BHG16}.  It consists of both a thick
\citep{We94, Bi97} and a thin bar component \citep{Ha00, We15}. The
thick component has a strong peanut shape seen in star counts
\citep{MZ10, WG13} and infrared photometry \citep{NL16}. Both gas
\citep{DHT01} and stars \citep{Ra09, Sa19} in the inner Galaxy show
strong non-circular streaming motions dominated by the bar
\citep{Fu99, Li16, Sh10, Po17}, and as in external boxy-peanut bulges,
the stellar velocity field shows cylindrical rotation \citep{Ho09}.

\citet{SW93} and \citet{BT08} have summarized much of the old
theoretical work that had led to considerable insight into the
structure of bars.  For the most part, this body of work considered
the bar to be a steadily rotating rigid object, and \citet{SS87} found
most of the expected orbit families in the frozen the potential of
their 2D $N$-body bar.  More recently, \citet{Va16} used orbits from
their simulations to demonstrate the relationships between the
traditional classification of orbits in bars with those of tri-axial
ellipsoids, while \citet{GLA16} found that the bar in their simulation
was sufficiently steady that they could fit a steadily rotating bar
model and apply the more modern apparatus of frequency analysis
developed by \citet{La90}.  \citet{Po15} and \citet{Ab17} also studied
the 3D shapes of particle orbits in the frozen potentials of their
$N$-body bars, focusing in both cases on orbits that supported the
peanut shape.

Many authors have reported simulations in which a bar thickened as a
result of a buckling instability \citep[\eg][]{Ra91, De05, MV06,
  Co20}.  Following \citet{CS81}, \citet{Co90} suggested that the
thickening mechanism was related to a resonance between the
frequencies of the periodic motions along and normal to the bar
mid-plane.  This more gradual mechanism was further developed by
\citet{Qu14}, who elegantly described how the vertical motions of
stars could be increased as the resonance swept past them.

In this paper, we study how the structure of a bar changes as it
evolves, an objective shared by \citet{PWK16, PWK19a, PWK19b}.  Their
papers concentrated on the halo distortion, the in-plane motion in the
bar, and torques, but our principal focus here is on the mechanisms of
bar thickening in the evolving potential of the simulated bar
\citep[see also][who focused on the buckling instability only]{Lo19}.
We present simulations that illustrate three thickening mechanisms:
one in which the bar thickened through a buckling instability, and a
second that seems to show that bar orbits were heated vertically by
the passage through the 2:1 vertical resonance, as described by
\citet{Qu14}.  We were surprised to find a third mechanism of gradual
trapping of orbits into the 2:1 resonance that has not previously been
identified in simulations, to our knowledge, but had been proposed by
\citet{Qu02}.  We found this behaviour in a somewhat slow bar,
\ie\ one in which corotation was 1.6 times the bar semi-major axis, a
larger than usual fraction \citep[\eg][]{Ag15}.

\begin{figure}
\includegraphics[width=\hsize,angle=0]{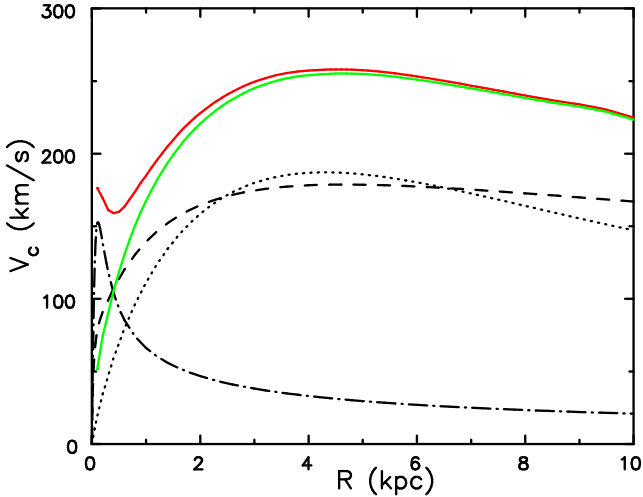}
\caption{The initial rotation curves of our models with (red) a
  nuclear star cluster and without (green).  These curves are
  determined from the particles.  The other lines indicate the
  separate contributions of the individual components: the disc
  (dotted), the nuclear star cluster (dot-dashed), and the Hernquist
  halo compressed, in this case, by both the disc and the nuclear star
  cluster (dashed).  The scaling from model units to physical units is
  described in the text.}
\label{fig.rotcur}
\end{figure}

\section{Technique}
\label{sec.methods}
We use collisionless simulations to study the processes of bar
thickening, and start from bar-unstable equilibrium disc models
embedded in spherical halos.  Our models for the disc and halo were
selected for convenience only, with no inention to match any
particular galaxy.

\subsection{Mass models}
\label{sec.models}
We create models of discs and halos, and in some cases add a nuclear
mass component, all composed of particles.  The disc is an exponential
with a Gaussian vertical density profile
\begin{equation}
\rho_d(R,z) = {M_d \over (2\pi)^{3/2} R_d^2z_0}e^{-(R/R_d + z^2/2z_0^2)}
\label{eq.disk}
\end{equation}
where $M_d$ is the disc mass, $R_d$ the scale length, and $z_0 =
0.1R_d$.  The disc surface density is tapered to zero using a cubic
polynomial over the radial range $4.5<R/R_d<5.0$.

We use an initially isotropic spherical Hernquist halo \citep{He90}
having the density profile
\begin{equation}
\rho_h(r) = {M_h \over 2\pi}{r_h^4 \over r(r+ r_h)^3},
\label{eq.halo}
\end{equation}
where $M_h=10M_d$ and $r_h=5R_d$.  We allow for the presence of the
additional components (the disc and perhaps a nuclear component) by
compressing the halo adiabatically by the method described by
\citet{SM05}.

In some models we added a nuclear star cluster modeled as an isotropic
Plummer sphere having a mass $M_d/40$ and a core radius of $R_d/20$.

The disc particles were given equilibrium orbital velocities with an
initial radial velocity dispersion, $\sigma_R$, so that $Q=1$
(neglecting corrections for disc thickness and gravity softening) at
all radii, where
\begin{equation}
Q = \sigma_R {\kappa \over 3.36 G\Sigma},
\label{eq.disp}
\end{equation}
$\Sigma(R)$ is the projected surface density, and $\kappa(R)$ is the
numerically-estimated epicyclic frequency in the disc mid-plane.  The
azimuthal dispersion, asymmetric drift, and vertical velocity
dispersion are all set by solving the Jeans equations \citep{BT08}.

Although our models were not intended to match the Milky Way, in which
the inner halo has a lower density \citep{Po17}, we nevertheless scale
them to the Milky Way.  Choosing $R_d=2\;$kpc and a dynamical time
$\tau_{\rm dyn} \equiv (R_d^3/GM_d)^{1/2} = 6.5\;$Myr yields bar sizes
that resemble that in the Milky Way, an orbital speed of $\sim
250\;$km~s$^{-1}$ at a radius of 8~kpc in the disc mid-plane, and a
disk scale height $z_0 = 200\;$pc.  These choices imply a mass unit,
the mass of the disc, $M_d \simeq 4.21 \times 10^{10}\;$\Msun, a unit
of velocity $R_d/\tau_{\rm dyn} \simeq 301\;$km~s$^{-1}$, and a
frequency unit of $\tau_{\rm dyn}^{-1} \simeq 47.7\;$\funit.

The initial rotation curves of our equilibrium models with (red) and
without (green) a nuclear star cluster are illustrated in
Figure~\ref{fig.rotcur}.

\begin{table}
\caption{Numerical parameters}
\label{tab.params}
\begin{tabular}{@{}ll}
Polar grid size & 85 $\times$ 128 $\times$ 125 \\
$z$ spacing & $0.02R_d=40\;$pc \\
Active sectoral harmonics & $0 \leq m \leq 8$ \\
Grid distance unit ($h_R$) & $0.1R_d = 200\;$pc \\
Softening length & $0.1R_d = 200\;$pc \\
Spherical grid size & 500 shells \\
Outer radius & $60R_d = 120\;$kpc \\
Active spherical harmonics & $0 \leq l \leq 4$ \\
Number of disc particles & $10^6$ \\
Number of halo particles & $10^6$ \\
Basic time-step & $0.01[R_d^3/(GM_d)]^{1/2} = 6.5\times10^4\;$yr \\
Time step zones & 5 \\
\end{tabular}
\end{table}

\subsection{Numerical method}
We evolve these models using the hybrid grid option of the {\tt
  GALAXY} code, which is fully described in the on-line manual
\citep{Se14}; the code itself is available for download.  The mutual
attractions of all particles are computed at every step.  To achieve
this, we employ a cylindrical polar 3D grid to compute the
gravitational field of the disc particles together with a spherical
grid for the halo and nuclear star cluster, if present.  For
efficiency, we use block time steps in a series of spatially-defined,
spherical zones, in which particles farther from the centre move on
time steps that are successively increased by factors of two for zones
more distant from the centre.

Table~\ref{tab.params} gives the values of the numerical parameters
for simulations lacking a nuclear mass component.  We shorten the
basic time step by a factor of 4 and include 2 extra time step zones
when a nuclear star cluster, represented by $2.5\times10^4$ particles
is present.  We have verified that our results are insensitive to
reasonable changes to these numerical parameters.

\section{Results}
\label{sec.results}
We computed the evolution of all models to $5.2\;$Gyr, during which
time they all formed strong bars.  The bars in models with a nuclear
mass component tended not to buckle, and instead puffed up gradually
over time, whereas those in models that lacked the dense central
component did buckle soon after their formation, but also continued to
puff up over time.  While we have computed many models, we present
just three in this section.

\begin{figure}
\includegraphics[width=\hsize,angle=0]{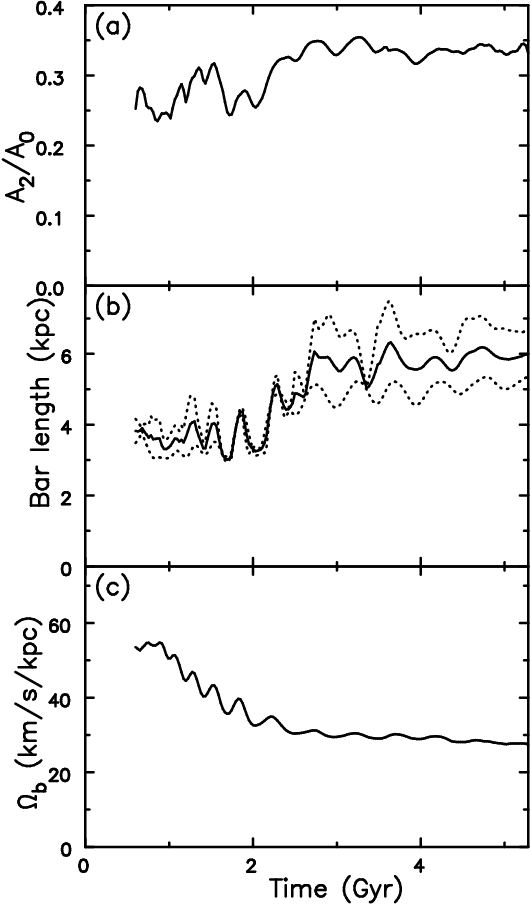}
\smallskip
\includegraphics[width=\hsize,angle=0]{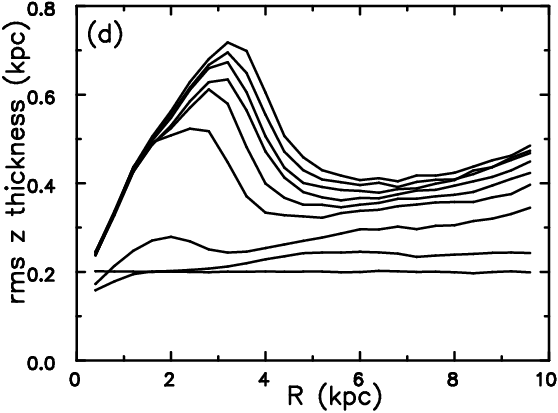}
\caption{Model A: The time evolution of (a) the bar amplitude, (b)
  the bar length (solid line) and (c) the bar pattern speed.  (d)
  Radial variation of the azimuthally averaged rms thickness of disc
  particles at intervals of $0.65\;$Gyr from $t=0$ to $t=5.2\;$Gyr.
  The thickness increases monotonically at most radii.}
\label{fig.4625_bar}
\end{figure}

\subsection{A bar that buckled}
\label{sec.modelA}
We first describe model A, which lacked a nuclear mass component.
The initial equilibrium disc was globally unstable to the formation
of a bar, which had developed by $t \sim 0.7\;$Gyr. The bar later
buckled in the usual way at $t \sim 1.5\;$Gyr.

Figure~\ref{fig.4625_bar} presents the time evolution of (a) the
amplitude, (b) semi-major axis $a_{\rm B}$, (c) pattern speed of the
bar, and (d) the radial variation of the rms $z$-thickness of the disc
particles at intervals of 0.65~Gyr.  The bar amplitude is $A_2/A_0(t)
= |\sum_j e^{2i\phi_j}|/N_d$, where $\phi_j(t)$ is the azimuth of the
$j$-th disc particle at time $t$ and $N_d$ is the number of disc
particles, which all have equal mass.  The bar length (panel b) is
the average of the two dotted curves, which were estimated by the
methods described in \citet{DS00}.  The initial semi-major axis is
$a_B\la4\;$kpc, but by $t=2.6\;$Gyr it had grown in length to $a_B
\approx 6\;$kpc.  The pattern speed (panel c) decreased rapidly to
$t\sim2\;$Gyr due to friction with the halo, which weakened over time.
At late times, the corotation radius $R_{\rm CR}\la 8\;$kpc so
that the dimensionless ratio ${\cal R} \equiv R_{\rm CR}/a_B \sim
1.3$.

\begin{figure}
\includegraphics[width=\hsize,angle=0]{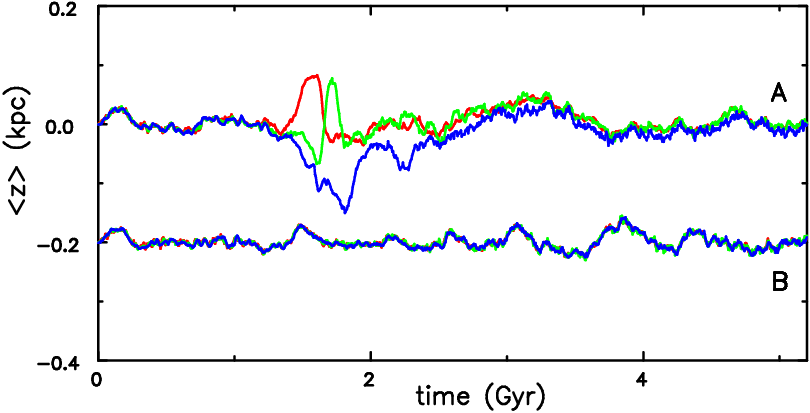}
\caption{The time evolution of the mean $z$-height of particles in
  three radial bins of width 400~pc centered on 1~kpc (red), 2.2~kpc
  (green) and 3.4~kpc (blue) in models A and B.  Values for
  model B have been shifted by $-200$~pc for clarity.}
\label{fig.zbend}
\end{figure}

Figure~\ref{fig.zbend} displays the time evolution of the mean
$z$-coordinate of the disc particles in three different radial bins
for this model and for model B.  The curves of different colours,
which indicate different radial bins, separate for the period $1.3 \la
t \la 2.5\;$Gyr indicating that the disc, which is dominated by the
bar at these radii, is flexing.

The vertical thickness of the bar (Figure~\ref{fig.4625_bar}d)
increased substantially between the third and fourth curves, which are
for $t=1.3$ and $t=1.95\;$Gyr respectively; this interval brackets the
time of the buckling instability.  The bar continued to puff up more
gradually until the last time shown ($t=5.2\;$Gyr).  The radial range
of the thickest part of the bar is $2 \la R \la 4\;$kpc, thus the
outermost $\sim 1/3$ of its length could be described as a ``thin
bar'', which had a thickness similar to that of the disc just beyond
the bar's end.

The bar also weakens temporarily at the time of buckling instability
(Figure~\ref{fig.4625_bar}a), as has been reported before
\citep[\eg][]{Ra91, MVS04, De05}.

\begin{figure}
\includegraphics[width=\hsize,angle=0]{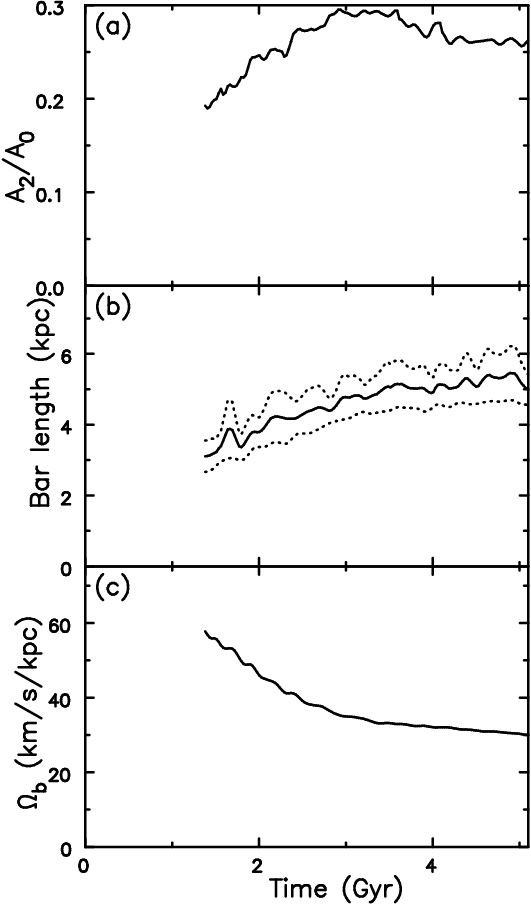}
\smallskip
\includegraphics[width=\hsize,angle=0]{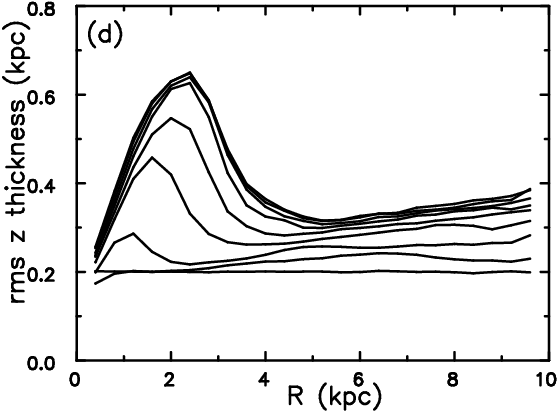}
\caption{Model B: The time evolution of (a) the bar amplitude, (b)
  the bar length (solid line) and (c) the bar pattern speed.  (d)
  Radial variation of the azimuthally averaged rms thickness of disc
  particles at intervals of $0.65\;$Gyr from $t=0$ to $t=5.2\;$Gyr.
  The thickness increases monotonically at most radii.}
\label{fig.4642_bar}
\end{figure}

\subsection{A bar that did not buckle}
\label{sec.modelB}
We included the nuclear star cluster described in \S\ref{sec.models}
in model B, which also formed a strong bar.  Linear stability theory
would predict that such a dense central mass should prevent bar
formation by inserting an inner Lindbald resonance to block the
feedback loop that drives the instability \citep{To81, BT08}, but
non-linear orbit trapping \citep[\eg][]{Se89} generally overwhelms the
resonance in most simulations, allowing a bar to form.

Figure~\ref{fig.4642_bar} shows that the bar took a little longer to
develop than in model A, but it was well established by $\sim2\;$Gyr
when it had already begun to thicken.  However, it did not appear to
buckle at any time, as shown in Figure~\ref{fig.zbend}.  Values of
$\langle z \rangle$ in model B fluctuated over time, but the blue
line almost perfectly overlays the red and green lines, indicating
that mid-plane moved equally at all three radii, and that the bar was
not flexing.  This contrasts with the behaviour in model A in the
same Figure, in which different radial ranges were displaced in
opposite senses for a period.

The initial bar in model B has $a_B \simeq 3\;$kpc
(Figure~\ref{fig.4642_bar}b), and it grows to $a_B \simeq 5.2\;$kpc by
the end.  Again the bar pattern speed (panel c) slowed rapidly at
first and then more gradually.  The time variation of the corotation
radius, $R_{\rm CR}$, and $a_{\rm B}$ conspired to make their
dimensionless ratio ${\cal R} \simeq 1.4$ for most of the evolution.

Panel (d) shows that by $t=1.95\;$Gyr (the fourth curve), the bar had
thickened over the range $1 \leq R \leq 2\;$kpc, and it continued to
thicken as it also grew in length.  As for model A, the outer bar ($4
\la R \la 5.2\;$kpc) is no thicker than the disc beyond the bar end at
later times, while the thickest part of the bar lies between $1 \la R
\la 3\;$kpc.

\begin{figure}
\includegraphics[width=.98\hsize,angle=0]{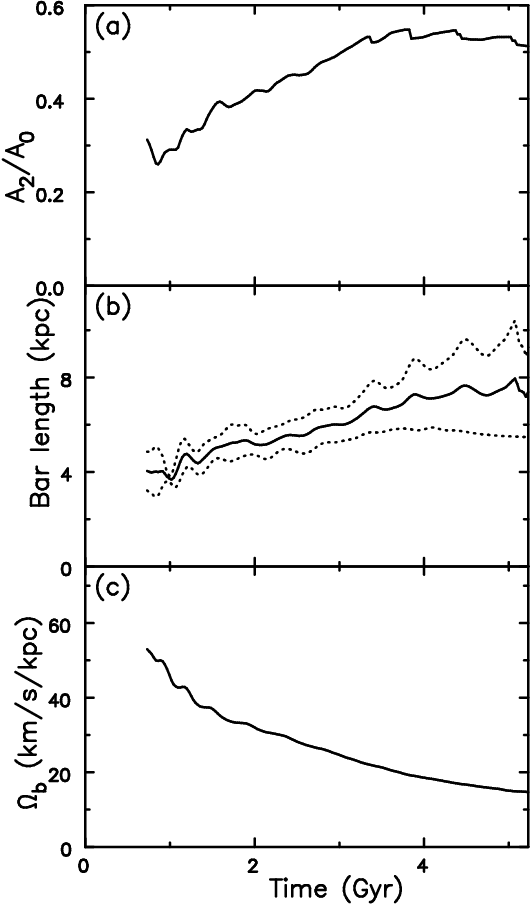}
\smallskip
\includegraphics[width=.98\hsize,angle=0]{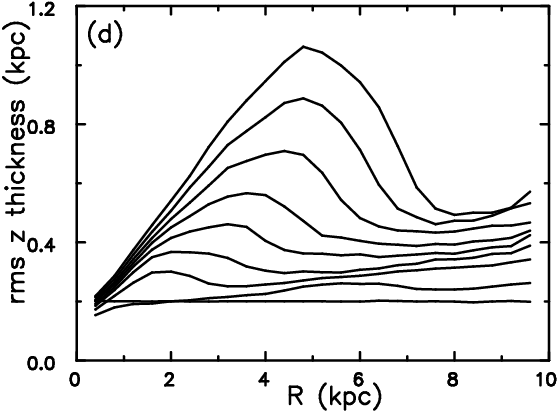}
\caption{Model C: The time evolution of (a) the bar amplitude, (b) the
  bar length (solid line) and (c) the bar pattern speed.  (d) Radial
  varition of the azimuthally averaged rms thickness of disc particles
  at intervals of $0.65\;$Gyr from $t=0$ to $t=5.2\;$Gyr.  The
  thickness increases monotonically at most radii.}
\label{fig.4645_bar}
\end{figure}

\subsection{A remarkable comparison}
\label{sec.modelC}
It is possible to inhibit the buckling instability by imposing
reflection symmetry about the mid-plane at each step of the
simulation, as reported by \citet{FP90}.  This is easily achieved with
a grid-code, such as {\tt GALAXY}, by replacing the masses assigned to
the grid points by the average of the values above and below the
mid-plane before determining the gravitational field.

We pursued this strategy in model C, which was identical in every
respect to model A except for the imposition of vertical reflection
symmetry.  As we intended, the buckling instability of the bar was
indeed inhibited, but the evolution of the bars differed from those in
model A to a surprising extent.

Because the buckling instability was suppressed, the bar amplitude
grew monotonically (Fig.~\ref{fig.4645_bar}a), but the final bar
amplitude in C was more than 50\% greater than in model A
(Fig.~\ref{fig.4625_bar}a).  The bar semi-axes started at the same
value, as expected, although the bar became longer when buckling was
supressed (panel b).  The increased bar amplitudes and lengths over
that in models A created a stronger quadrupole moment that caused
stronger friction with the halo, slowing the bar to a greater extent
(panel c), and causing the dimensionless ratio to rise to ${\cal R}
\ga 1.6$.  In fact, the disc component of model C had lost over 30\%
of its initial angular momentum to the halo by the end of the run,
which is twice that lost in model A.

But the most surprising consequence of inhibiting buckling is that the
vertical thickness of the bar in C continued to grow throughout the
evolution, reaching much greater values (Fig.~\ref{fig.4645_bar}d)
than in model A (Fig.~\ref{fig.4625_bar}d); \ie\ suppressing
buckling resulted much thicker final bar!  Furthermore, the bar
thickened over almost its entire extent, save for the very centre, and
there was little if any outer thin bar.

\begin{figure*}
\includegraphics[width=.31\hsize,angle=0]{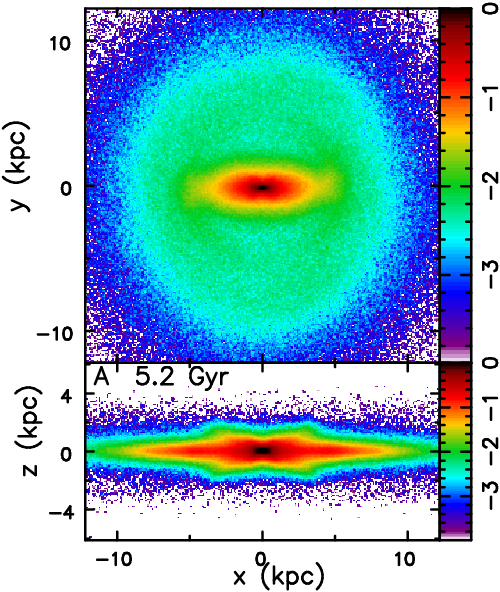}\hspace{0.5cm}
\includegraphics[width=.31\hsize,angle=0]{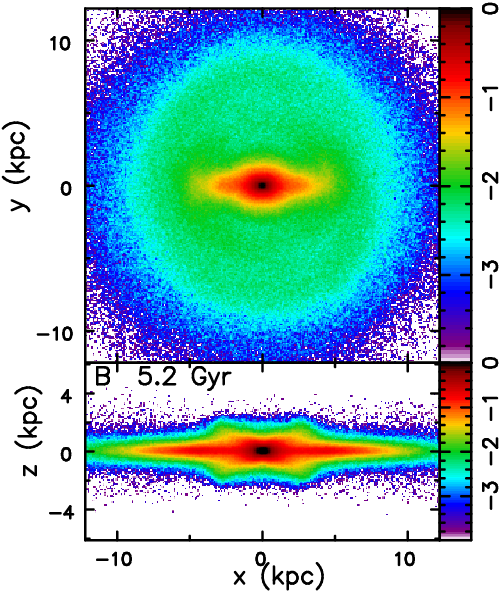}\hspace{0.5cm}
\includegraphics[width=.31\hsize,angle=0]{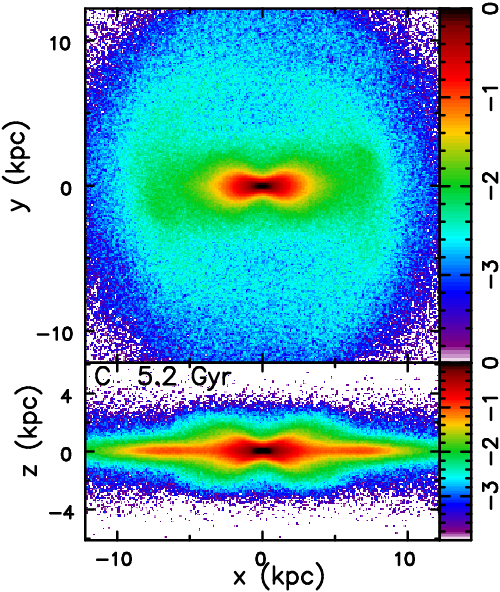}
\caption{Snapshots of models A (left), B (middle), and C (right) at
  the final time.  The logarithmic colour scale indicates projected
  density of the disc particles only, normalized to its peak value in
  each panel.  Each model has been rotated so that the bar is
  horizontal, and the lower panels show the edge-on view.}
\label{fig.project}
\end{figure*}

\subsection{Visual comparison}
\label{sec.vis}
Figure \ref{fig.project} presents snapshots of models A, B, and C at
the final time.  Note that the shapes of the bars in the pole-on
projection are all quite different.  The bar in model A is oval and
that in model B is more lens-shaped with a near axisymmetric centre.
The high density inner bar in model C has a more butterfly shape,
while the low-density outer bar is again oval.  The vertical
thicknesses all differ, as already reported, and it can be seen that
the greater thickness of the bar in model C largely stems from a
low-density envelope around the high density in the mid-plane.

\section{Analysis}
It is well known that bars become vertically thicker as a result of
the buckling instability \citep{Ra91, SM94}.  \citet{Co90} argued for
an alternative mechanism involving the 2:1 vertical resonance, at
which $\Omega_z = 2 \Omega_x$; here $\Omega_z$ is the vertical
freqency and $\Omega_x$ is the frequency of oscillation along the bar
major axis in the rotating bar frame.  \citet{Qu14} developed this
second idea and found that orbits could be elevated by becoming
trapped in the resonance, and argued that the resonance sweeps
outwards as the bar evolves, allowing orbits to be later released from
the resonance with an increased vertical oscillation amplitude.
\citet{Qu14} claimed evidence in support of their thickening mechanism
from analysis of a few snapshots taken from archival simulations.
Here we mount a much more robust test of their proposed mechanism.

The bars in our models evolve continuously in pattern speed, length,
and thickness.  Since our objective is to understand the mechanisms
that cause the thickness to change, we need to follow orbits in the
evolving model.

\begin{figure}
\includegraphics[width=\hsize,angle=0]{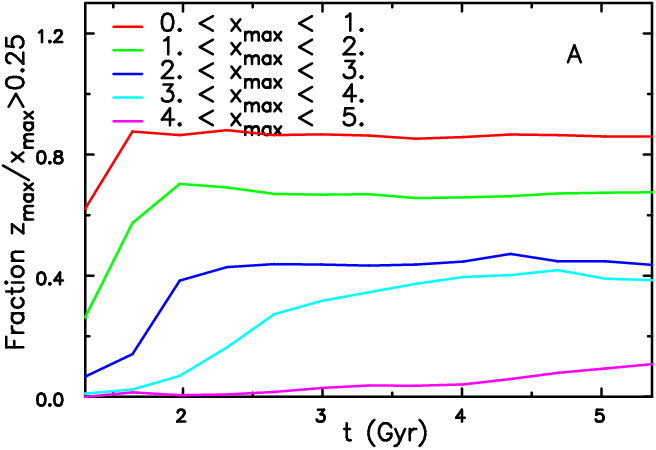}
\includegraphics[width=\hsize,angle=0]{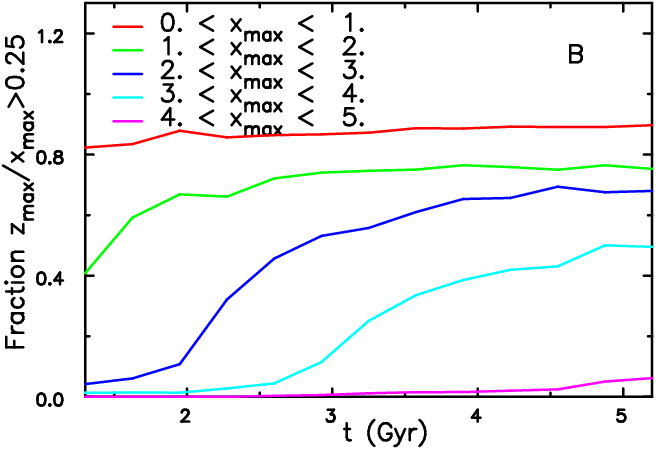}
\includegraphics[width=\hsize,angle=0]{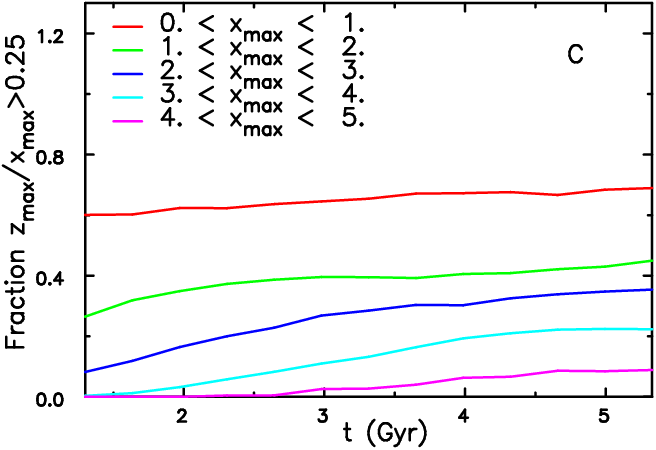}
\caption{The time evolution of the fraction of orbits in all three
  models that have $z_{\rm max}/x_{\rm max}>0.25$.  The lines are
  coloured by the extent, in kpc, of the orbits along the bar axis.}
\label{fig.hifracts}
\end{figure}

\subsection{Orbits in an evolving bar}
\label{sec.orbits}
Accordingly, we recorded the position coordinates of a randomly
selected fraction of the disc particles at intervals of $0.16\tau_{\rm
  dyn} \simeq 1\;$Myr. The fraction was 1\% in models A and C, and 2\%
in model B.  We also measured the position angle of the bar major axis
over time from the mass-weighted average of the $m=2$ sectoral
harmonic over the radial range $0.4 \leq R \leq 8\;$kpc.  For each
recorded orbit, we rotated the position about the $z$-axis at each
instant to a frame in which the bar position angle is horizontal.
Unless otherwaise stated, we selected orbits for which $|x|$ was
always $<6\;$kpc for models A and B, and $|x|<8\;$kpc in model C, a
criterion that eliminated slightly fewer than half the orbits, which
were those that strayed from, or never were in, the bar.

For every orbit, we identified the turning points in each of the
separate coordinates $(x,y,z)$ in the bar frame, and computed moving
averages of the absolute values at ten successive extrema that bracket
a given time, \ie\ over five full periods, which we denote $x_{\rm
  max}(t)$, \etc

Figure~\ref{fig.hifracts} plots the time evolution of the fraction of
the bar particles in all three models whose greatest vertical
departures from the mid-plane $z_{\rm max}/x_{\rm max}>0.25$ at each
moment.  We separate them by their $x_{\rm max}$ values, in the ranges
indicated by the colours.  Orbits confined to the inner bar (red line)
are nearly all thick, by this criterion, while those that extend into
the outer bar (magenta line) are almost all thin, although both lines
rise gently with time.  For the intermediate ranges, indicated by the
green, blue, and cyan lines, the behaviour differs in all three
models.  In model A, the red, green and blue lines all rise until the
buckling instability saturates, while heating in the outer bar (cyan
line) is more gradual.  The heated fraction also increases
substantially in model B, with smaller orbits rising more rapidly
early on; both the green and blue lines suggest that $>60$\% of the
orbits of intermediate values of $x_{\rm max}$ have large $z_{\rm
  max}/x_{\rm max}$ by the end of the evolution.  The heated fractions
in model C are generally lower than those in models A and B, which may
seem odd, given that the rms thickness was greatest in this bar.
However, as noted above, edge-on views in (Fig.~\ref{fig.project})
reveal that the greater rms thickness in model C is due to low-density
halo around the bar, consistent with a smaller heated fraction. (We
provide further supporting evidence for this statement in \S5.2
below.)

\begin{figure*}
\includegraphics[width=\hsize,angle=0]{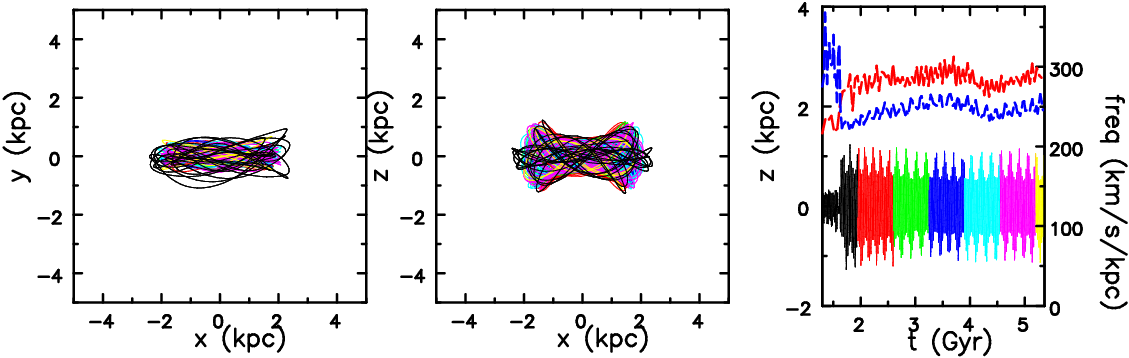}
\includegraphics[width=\hsize,angle=0]{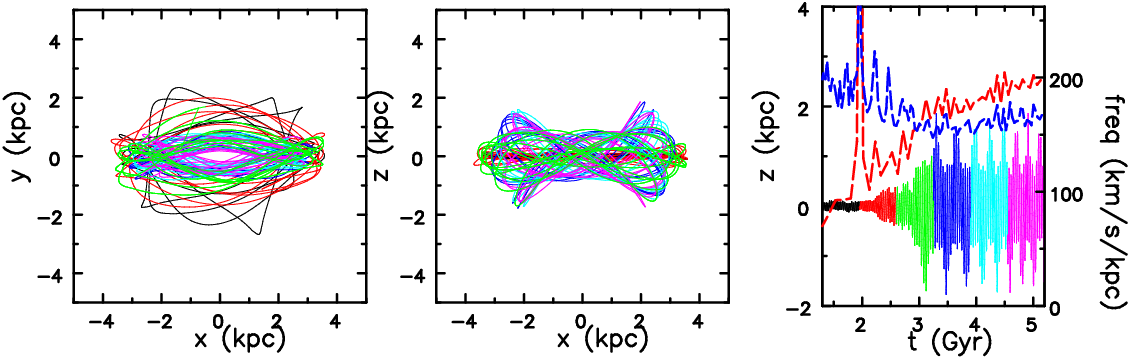}
\caption{Good examples of vertical heating in models A (above) and B
  (below).  The left and middle panels show $(x,y)$ and $(x,z)$
  projections of the orbit in a frame that is oriented at each moment
  with the bar major axis, which is horizontal.  The solid line in the
  right panel records the time evolution of the $z$ coordinate from
  $t=1.3\;$Gyr; the colour of this line changes every 0.65~Gyr, as it
  does in the other two panels, although the black (green) part is
  plotted last in model A (B).  The broken lines in the right panels
  indicate crude estimates of $2\tilde\Omega_x$ (red dashed) and
  $\tilde\Omega_z$ (blue dashed).  In model A (above), the principal
  increase in the amplitude of the $z$ oscillation occurs at
  $t\sim1.7\;$Gyr, which is about when the bar buckles.  The frequency
  changes in model B are more gradual, and heating occurs around the
  time that the two frequency curves cross, when $2\tilde\Omega_x
  \simeq \tilde\Omega_z$.  The subsequent semi-periodic modulation of
  the vertical oscillation amplitude of both orbits is caused by the
  turning points of the vertical motion varying with the $x$ position
  as illustrated in the middle panels.}
\label{fig.orbit1286}
\end{figure*}

\subsection{Orbit frequencies}
Using our recorded orbits in the rotating bar frame, we determined the
times at which the motion reversed in each coordinate, and define a
rough frequency to be $\tilde\Omega_i \approx 2\pi/\tilde\tau_i$,
where the period, $\tilde\tau_i$, is the time between two successive
maxima of the particle's $i$\/th coordinate, \ie\ a full period.  Here
we use a tilde to distinguish our numerically estimated values from
quantities that could be more meaningfully defined through
action-angle variables, say.  Good examples of orbits from models A
and B are illustrated in Figure~\ref{fig.orbit1286}; the two
projections in the left and centre panels are drawn in the frame that
is rotated at each instant so that the bar major axis is horizontal.
The colour of the line in each frame changes with time, as in the
right hand panel, although the black (green) part is plotted last in
order to show the orbit shape during the period of vertical heating
more clearly.

The orbit in model A (above) is heated abruptly as the buckling
instability saturates when $\tilde\Omega_z$, dashed blue curve in the
right panel, decreases.  Notice that $\tilde\Omega_x$ also rises at
the same moment, because the horizonal extent of the orbit decreases
(middle panel).

The $x-y$ projection of the orbit in model B (lower left panel) is
elongated parallel to the bar and never approaches the centre, but it
does become noticeably skinnier over time.  This appearance suggests
it is trapped about an $x_1$ periodic orbit \citep[see \eg][]{SW93}
throughout.  This in itself is remarkable, since the bar pattern speed
decreases by a factor of 2, its amplitude increases by $\ga 30$\%, the
thickness of the inner bar more than doubles, $\tilde\Omega_z$
decreases by $\sim25$\%, $\tilde\Omega_x$ more than doubles, and the
$z$-oscillation amplitude increases 10-fold over the period
illustrated, yet the basic character of the orbit in its $x-y$
projection scarcely changes!  This robust property of the orbit is far
from unique -- we have noticed similar behaviour in many other orbits.

Our crude frequency estimates are shown by the broken curves in the
right panel.\footnote{Anomalous frequency estimates, such as at $t
  \simeq 2\;$Gyr in this Figure, can occur when sign changes of $\dot
  x$ or $\dot z$ arise from a more complicated orbit shape.}  Even in
a steady potential, the motion of regular orbits \citep{SW93, BT08}
can be decomposed into librations about an underlying, or parent,
periodic orbit.  The fundamental frequencies of the parent periodic
orbit will generally be incommensurable with the libration
frequencies, causing oscillations in all three principal axes to be
aperiodic, which is reflected in the short-term variation of the
estimated frequencies shown in this Figure.  

The frequencies of this orbit also change on a longer time scale as
the bar evolves; $\Omega_z$ decreases because the bar thickens,
reducing the density, while the increase in $\Omega_x$ results from
the more rapid apparent motion of the particle as the rotating frame
of the bar slows.  While there is substantial jitter in the individual
frequency estimates, the two dashed curves clearly cross near
$t=3\;$Gyr, which is about the time of a rapid increase in the
vertical oscillation amplitude of this orbit, consistent with the
mechanism proposed by \citet{Qu14}.  We will show in
\S\ref{sec.Q14test} below that this behaviour is quite general.

The modulation of the $z$ excursions at later times in
Figure~\ref{fig.orbit1286} is another property that occurs in many
other orbits, and is accounted for as follows.  As may be seen from
the $(x,z)$ projection of the orbit in the middle panel, the
absolutely largest values of $|z|$ occur when $|x|\ga 2\;$kpc.  But
$z$ maxima are also reached when $|x| \la 1.6\;$kpc, where the
envelope of the orbit in the middle panel has a flattish waist, which
must be caused by a stronger vertical restoring force where the mass
density is higher.  Since $\tilde\Omega_x$ and $\tilde\Omega_z$ are
generally incommensurable, the vertical oscillation amplitude is
modulated as shown, and the increasing difference between the $x$- and
$z$-oscillation frequencies after $t\sim3\;$Gyr results in the
observed decreasing modulation period of the $z$-oscillation
amplitude.  Note that the oscillation amplitude variations appear to
have little effect on the estimated vertical frequency (dashed blue
curve in the right panel).

Once again, in order to obtain more smoothly varying frequency
estimates, we henceforth compute each frequency from moving averages
over 10 full periods in each coordinate.

\subsection{Collective support for buckling}
\label{sec.buckle}
Particles confined within a bending bar experience a vertical driving
acceleration as a result of their horizontal motion along the bar.
When the vertical distortion is a single arch, the vertical forcing
frequency is twice that of its horizontal frequency along the bar.  If
this forcing frequency, $2\Omega_x$, is less than the natural
frequency of free oscillation in the vertical direction, $\Omega_z$,
then the particle will be able to follow the bend.  Any orbit for
which $\Omega_z < 2\Omega_x$ would respond out of phase with the bend
and would not cooperate with the bending instability.  Thus
\citet{MS94} argued that a supporting response to a buckling bar,
\ie\ a collective instability, is expected only for orbits that have
$\Omega_z > 2\Omega_x$.

\begin{figure}
\includegraphics[width=\hsize,angle=0]{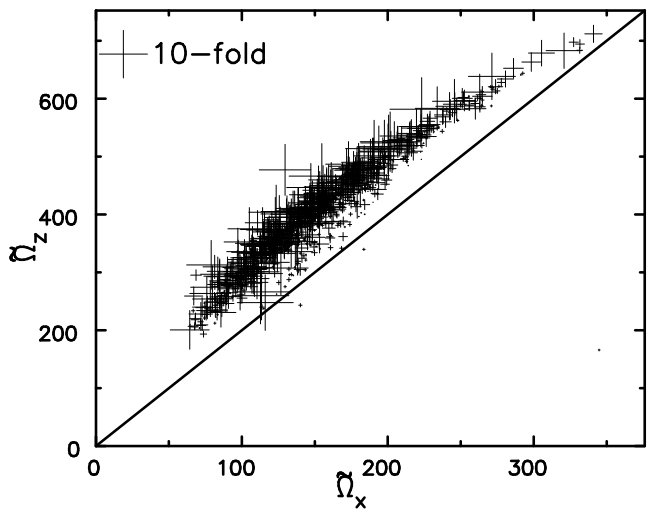}
\caption{Frequencies, estimated at $t=1.3\;$Gyr, of orbits in model A
  whose vertical excursions more than doubled at the time the bar
  buckled. The sizes of the symbols indicate the logarithm of half the
  factor by which $z_{\rm max}$ has grown over the interval $1.3 \pm
  0.3\;$Gyr.  The line has slope 2 and frequencies are in units of
  \funit.}
\label{fig.resbuk}
\end{figure}

\begin{figure*}
\includegraphics[width=\hsize,angle=0]{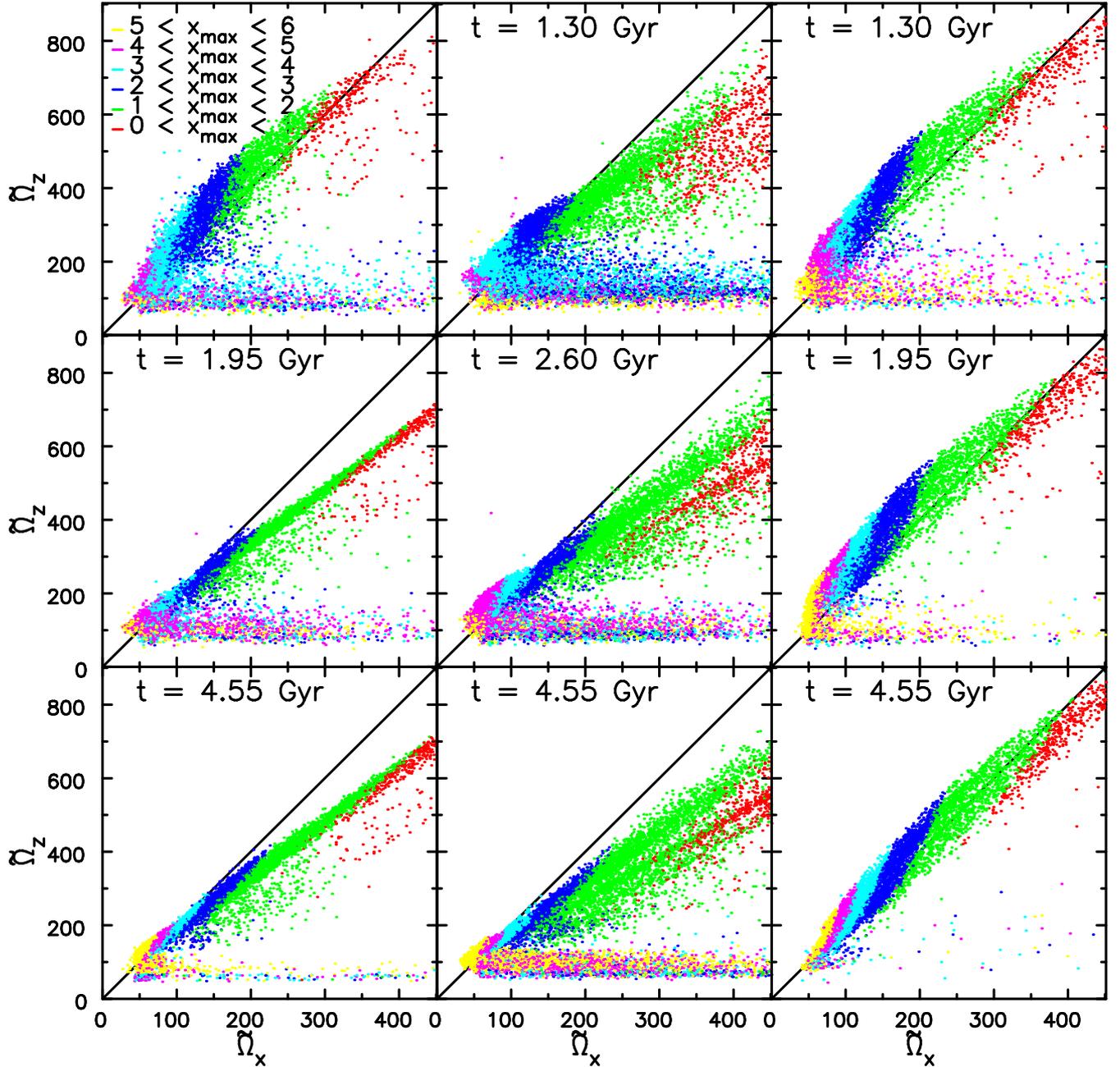}
\caption{The frequency distribution of all recorded orbits, except
  those in the far outer disc, in models A (left), B (middle), and C
  (right) soon after the bar formed (top row), towards the end of the
  simulation (bottom row), and at an intermediate time (middle row).
  The diagonal line marks the 2:1 resonance.  The approximately
  horizontal distributions of particles at low $\tilde\Omega_z$ are
  mostly outside the bar.  The range of $x_{\rm max}$ values in kpc
  are indicated by the colours.}
\label{fig.frqplt}
\end{figure*}

Figure~\ref{fig.resbuk} records the frequencies of orbits in model A
for which the magnitude of the vertical excursion more than doubled at
the time of the buckling event $t \approx 1.5\;$Gyr.  We also selected
orbits that were significantly elongated in the direction of the bar,
such that $y_{\rm max} < 0.3x_{\rm max}$.  The frequencies
$\tilde\Omega_x$ and $\tilde\Omega_z$ were estimated at $t=1.3\;$Gyr,
\ie\ in the early stages of the bending instability.  Almost all these
orbits have $\tilde\Omega_z > 2\tilde\Omega_x$, which is consistent
with their need to follow the growing bend of the bar.  It is very
likely that our frequency estimates were incorrect for the three
points that are below the line.

The left-hand panels of Figure~\ref{fig.frqplt} report the frequencies
of all recorded orbits (see \S\ref{sec.orbits}), excluding only those
in the far outer disc, at three different times in model A, and are
color coded by ranges of $x_{\rm max}$.  The top panel is for the same
time as Figure~\ref{fig.resbuk} but now includes orbits lying below
the resonance line whose vertical oscillation amplitude did not
increase over the period of the buckling instability.  Particles in
the disc outside the bar all have low values of $\tilde\Omega_z$.
Recall that $\tilde\Omega_x$ is estimated from intervals between $\dot
x$ reversals; eccentric orbits outside the bar will have irregular
shapes when viewed in the bar frame, causing a broad spread in this
plotted frequency that has little physical significance.

The middle and bottom left panels of Figure~\ref{fig.frqplt} show the
same quantities later in the evolution, at $t=1.95\,$Gyr and
$t=4.55\,$Gyr respectively, of the same simulation.  It is remarkable
that the majority of bar orbits at the last time are tightly
distributed about a line that has slope $\sim1.38$ and a non-zero
intercept and this arrangement was already in place immediately after
the buckling instability had run its course (at $t=1.95\,$Gyr).  The
vertical frequencies of the orbits have decreased, especially in the
inner bar, because it has puffed up, but the fact that the global
properties of the bar at the later times are such that there is a
tight linear relationship between the vertical and horizontal
frequencies of almost all these orbits seems highly significant.
\citet{Po15} reported the distribution of frequency ratios of
particles in their frozen bars had a narrow and non-uniform
distribution over the range $1.5 \la \tilde\Omega_z/\tilde\Omega_x \la
2$ (in our notation) and a similar result was reported by \citet[][her
  Fig 12]{Lo19}, although the fuzzier line in her case seemed to pass
through the origin.  The non-zero intercept of the linear feature in
the bottom left panel of our Figure~\ref{fig.frqplt} and closest
rational slope of 11/8 makes a resonance explanation seem unpromising;
the feature deserves a follow-up study.

\begin{figure}
\includegraphics[width=\hsize,angle=0]{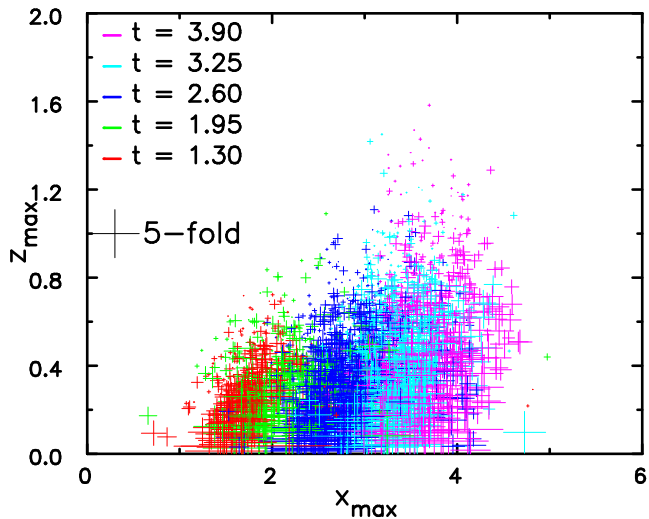}
\caption{The locations of nearly resonant skinny orbits in the plane
  of $(x_{\rm max},z_{\rm max})$ at five different times in model
  B. The sizes of the symbols indicate the logarithm of the factor by
  which $z_{\rm max}$ has grown over the interval $t\pm0.325\;$Gyr.
  Distances are in kpc, times in Gyr.}
\label{fig.resvst}
\end{figure}

\subsection{Gradual vertical heating}
\label{sec.Q14test}
As noted above, \citet{Qu14} suggested that orbits could also be
elevated by becoming trapped in a 2:1 vertical resonance that sweeps
outwards as the bar evolves, allowing orbits to be later released from
the resonance with an increased vertical oscillation amplitude.  The
bar in model B thickened without buckling, and we here examine whether
their puffing mechanism was indeed responsible for its vertical
thickening.

\begin{figure*}
\includegraphics[width=.338\hsize,angle=0]{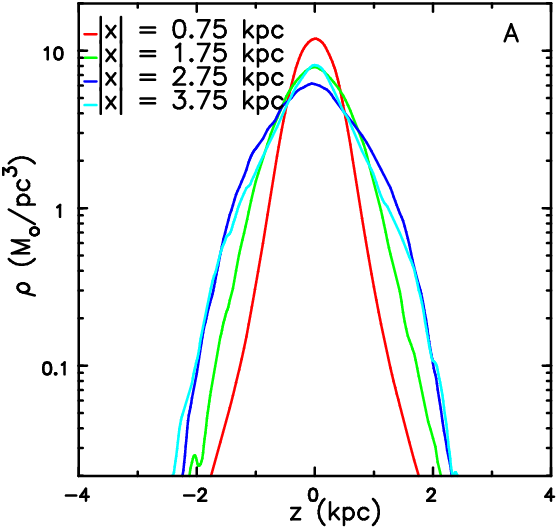}
\includegraphics[width=.3\hsize,angle=0]{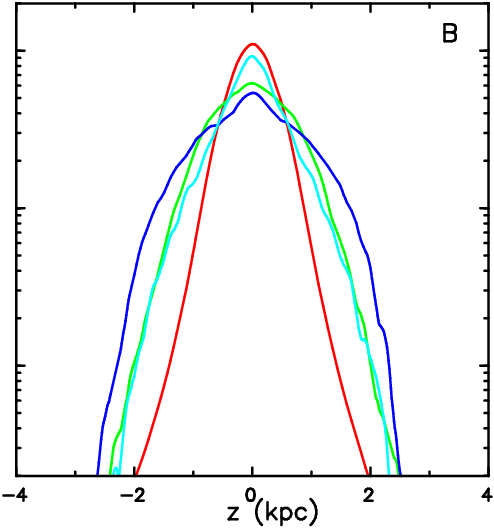}
\includegraphics[width=.3\hsize,angle=0]{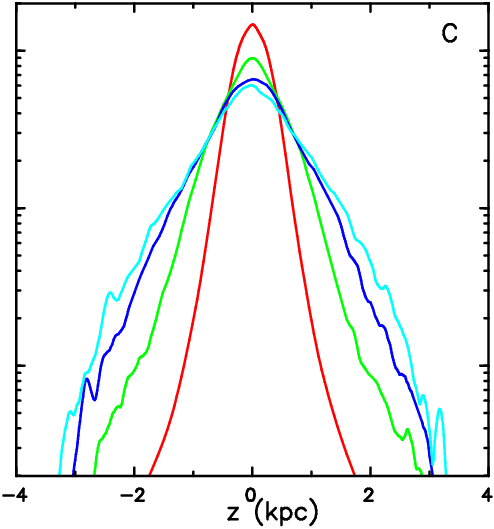}
\caption{The vertical density profiles at several distances along the
  bars at the final time.  The particles were binned in $\delta x =
  0.5\;$kpc and only those within $1.6\;$kpc of the bar major axis
  were included.  Notice that while the outer bar in model C has
  indeed become the thickest, as already noted, the mid-plane density
  of the inner bar is also the highest.}
\label{fig.rhozprof}
\end{figure*}

\begin{figure*}
\includegraphics[width=\hsize,angle=0]{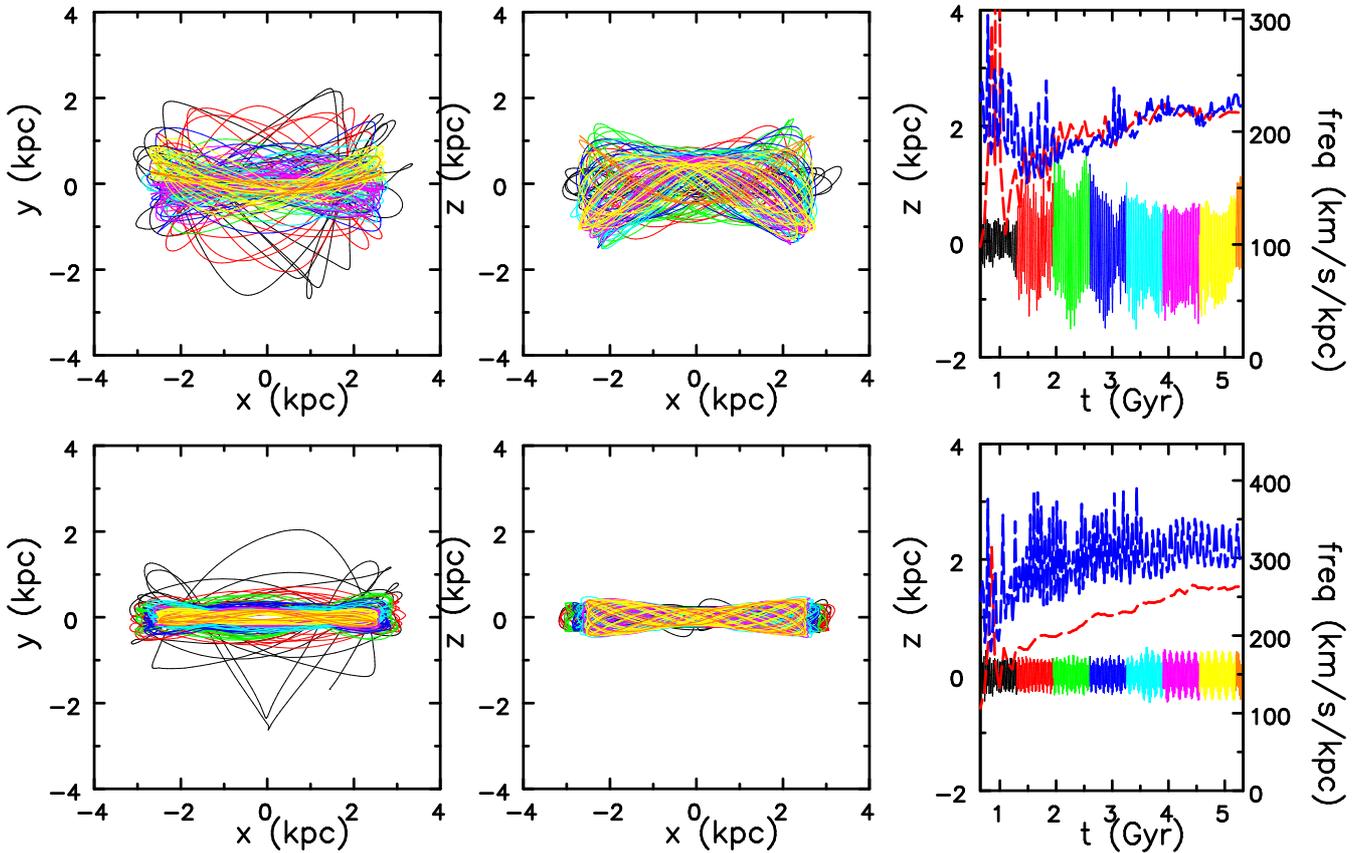}
\caption{Two orbits in model C.  The description of the panels is as
  in Figure~\ref{fig.orbit1286}. The orbit in the top row becomes
  trapped in the vertical resonance, while that is bottom row is not.}
\label{fig.orbit71}
\end{figure*}

In Figure~\ref{fig.resvst}, we plot the values of $z_{\rm max}$
against $x_{\rm max}$ of orbits in model B that are close to the
resonance, \ie\ those for which $|2\tilde\Omega_x - \tilde\Omega_z| <
4.8\;$\funit.  In this Figure, we have included only particles having
$y_{\rm max} < 0.3 x_{\rm max}$, \ie\ those whose orbits are highly
elongated parallel to the bar.  The colours indicate the times at
which all these quantities were measured and the sizes of the symbols
are proportional to the logarithm of the factor by which $z_{\rm max}$
has grown over an interval of $0.65\;$Gyr that brackets the time of
the measurement.  The orbits with the larger symbols have been lifted
to larger $z_{\rm max}$ by passage through the resonance.  The
resonance is broad in both $x_{\rm max}$ and $z_{\rm max}$ at any one
time, but there is a clear trend that the resonance moves outwards,
\ie\ to larger $x_{\rm max}$, over time as the bar slows.  The
anharmonic vertical potential causes the colour bands in
Figure~\ref{fig.resvst} to have a positive inclination; this follows
because $\Omega_z$ decreases with increasing $z_{\rm max}$ at fixed
$x_{\rm max}$, which requires a lower $\Omega_x$ and thus larger
$x_{\rm max}$ to be in resonance.

The symbols of each colour in Figure~\ref{fig.resvst} fill quite broad
and overlapping bands, suggesting that the resonance might not be
sharp.  There are several possible reasons that could cause a real or
apparent broadening of the resonance.  First, the approximate nature
of our frequency estimates will blur the resonance somewhat, although
particles that are wrongly identified as being in resonance will not
be trapped and therefore should not be elevated, giving rise to small
symbols in Figure~\ref{fig.resvst} that are outliers of the
distribution of truly resonant orbits of a given colour, as appears to
be the case.  Second, the time resolution of the color coding is quite
coarse and the bar will have evolved over the interval spanned by each
colour, causing a true spread in the location of the resonance.
Third, orbits having the same values of $(x_{\rm max},z_{\rm max})$ at
a given time could have differing shapes: some will be skinny,
\ie\ have small $y_{\rm max}$, while others will be rounder, and
whether they are in resonance will depend somewhat on their shape.  To
test this idea, we examined resonant orbits for which $y_{\rm max} >
0.3 x_{\rm max}$, \ie\ rounder orbits in the bar frame, finding that
the rounder orbits had a distribution in Figure~\ref{fig.resvst} that
was about twice as broad as the skinnier orbits, confirming this
hypothesis.  A fourth reason could be that the resonance is
intrinsically quite broad because of the large potential perturbation
caused by the strong bar.

The bar in model A continued to thicken after the buckling event
(Figure~\ref{fig.4625_bar}d).  We have similar evidence from this case
to that in model B (Figure~\ref{fig.resvst}), but do not present it,
that the secular thickening in this bar happened by the \citet{Qu14}
mechanism, since the orbits that were nearly resonant
$|2\tilde\Omega_x - \tilde\Omega_z| < 4.8\;$\funit\ were raised by the
largest factors, and again we observed the resonance moving gradually
out along the bar.

The middle panels of Figure~\ref{fig.frqplt} show the frequencies of
all orbits in model B at $t=1.3\,$Gyr (top), $t=2.6\,$Gyr (middle),
and $t=4.55\,$Gyr (bottom).  The addition of the nuclear star cluster
in model B should have raised the values of both frequencies in the
inner bar, shifting the points up and to the right in comparison with
model A, although the surface density in the inner bar in B had not
risen as much as in A because the bar was not quite fully formed at
this time.  However, it is clear that few of the red and green points
in model B, particles within 2~kpc of the centre, lie above the
diagonal at the earlier time.  This is the reason that the bar did not
buckle (see the discussion in \S\ref{sec.buckle}). When the bar had
settled and begun to puff up ($t=2.6\;$Gyr), the high frequency
orbits, in the inner bar, lie farther below the diagonal line, as for
model A, and in both runs one can see that those orbits that remain in
resonance at the later time are low-frequency orbits in the outer bar,
confirming in a different way that the resonance has swept outwards
along the bar to lower frequencies.

It is instructive to compare the evolution of the thicker fraction in
Figure~\ref{fig.hifracts}(B) with the changes between the top and bottom
panels for the same model B in Figure~\ref{fig.frqplt}.  (Note that
the colour codes for $x_{\rm max}$ are the same in both figures.)  The
fraction of particles that are heated to larger $z_{\rm max}$ rises
later in the evolution at larger radii, and these particles can be
seen to lie below the resonance line in the bottom middle panel of
Figure~\ref{fig.frqplt}; the only orbits still in resonance at this
time are in the outer bar ($x_{\rm max} \ga 3$).  Thus, we consider
this evidence, combined with that in Figure~\ref{fig.resvst}, to
indicate clearly that the bar in model B thickened as the resonance
swept out along the bar \citep{Qu14}.

The distribution of particles in frequency space in bottom middle
panel of Figure~\ref{fig.frqplt} (model B), has bifurcated into two
features: a broader version of the line in model A and a sharp line
having a shallower slope $\sim 1$ and a non-zero intercept.  Perhaps
even a third line having a yet shallower slope is faintly discernible.
Once again, we defer this unexpected finding to a follow-up
investigation.

\subsection{Trapping into resonance}
The behaviour in model C is different from that in both models A and
B, as may be seen in the right-hand panels of Figure~\ref{fig.frqplt}.
The similarity of the diagrams for models A and C at $t=1.3\;$Gyr is
because model C differed from model A only by the imposition of
vertical symmetry, which inhibited the buckling instability that was
on-going in model A at this time.  But in the middle and bottom panels
almost all particles in model C have remained close to the resonance
line, whereas the distribution in model A (bottom left) is narrower
and has a different slope and intercept.  Thus it seemed possible that
many bar particles in model C had been trapped into the 2:1 vertical
resonance.  Indeed, \citet{Qu02} had proposed this as a mechanism for
bar thickening, but we are unaware of any previously reported
simulation that supported her suggestion.

It may seem puzzling that the $\tilde\Omega_z$ values in the inner bar
are {\em higher} at the later time than in the other bars, even though
this bar thickened more.  However, we noted above that the disc in
model C gave up over 30\% of its angular momentum to the halo, mostly
from the inner disc, which allowed particles to settle closer to the
centre thereby creating a higher density in the bar than in model A.
This is illustrated in Figure~\ref{fig.rhozprof}, which compares the
vertical density profiles of all three bars at a range of
$x$-distances from the bar centre.  It can be seen that the mid-plane
density of the inner bar in model C is the highest of the three bars.
Thus the vertical restoring forces in the inner bar of model C are
stronger than those in model A, causing higher vertical frequencies.

The orbit presented in the top row of Figure~\ref{fig.orbit71} is a
good example of orbit trapping in model C, illustrating that once an
orbit gets into approximate resonance, it remains there for the
duration of the simulation.  That in the bottom row is not trapped by
the end of the simulation.  The later gradual rise of both
frequencies, visible in the rightmost panels, is due to the increasing
density in the inner bar as the model evolves.  Furthermore, the
vertical oscillation amplitude of the trapped orbit does not increase
monotonically, but fluctuates about a mean that is generally much
greater than its initial vertical amplitude, whereas the orbit that is
not trapped is not heated.  There are many orbits of both types in the
bar of model C.

The colour bands at the intermediate and later times in the right hand
panels of Figure~\ref{fig.frqplt} are quite sharp because the range of
$x_{\rm max}$ largely determines the horizontal frequency of the orbit
in the bar potential.  The boundaries between the colours have
positive slope, indicating that those with higher vertical frequency
or equivalently smaller vertical excursions, have slightly higher
horizontal frequencies at the same $x_{\rm max}$.

The distribution about the diagonal in the bottom, right panel of
Figure~\ref{fig.frqplt} is much broader than that of the line in the
bottom left panel, suggesting that the width is mostly intrinsic, and
not simply measuring errors in the frequencies, a point that is
supported by the sharpness of the boundaries of the colour bands.
However, not all the particles are resonant, as was illustrated by the
two examples in Figure~\ref{fig.orbit71}, and the non-resonant
particles, which have not been heated, are generally above the
diagonal. Orbits are heated as they become trapped, which causes
$\tilde\Omega_z$ to decrease.

Notice that there are very few green or blue points far below the
diagonal in the bottom right panel of Fig.~\ref{fig.frqplt}, implying
that most heated orbits remain trapped.  We have found that $\sim
10$\% of orbits that were heated later have persisting frequencies
below the resonance condition, suggesting they may have escaped from
the resonance.  However, the small frequency differences in some cases
make it difficult to be certain that all such orbits have fully
escaped.

It should be emphasized that model C is artificial, because we imposed
reflection symmetry about the mid-plane throughout its evolution.  It
is therefore possible that this third mechanism for bar thickening may
arise solely in simulations having this unphysical constraint.

\section{Discussion of the thickening mechanisms}
\label{sec.distinct}
The distributions of points in all three panels in the top row of
Figure~\ref{fig.frqplt} include many particles lying close to the
resonance line before the bar thickened.  It may therefore seem that
the resonance played a role in thickening all three bars and that the
diverging behaviour of each of the three models reflected different
consequences of the same underlying mechanism.

In order to show that this is not the case, we present two additional
simulations in which the initial disc thickness was halved to
$z_0=100\;$pc, but which were otherwise identical to models A and C.
We denote these models as \Ab\ and \Cb, respectively with the
subscript indicating the vertical scale of the initial disc.  Doubling
the disc density in this way strengthens the gravitational attraction
both vertically towards the mid-plane and in the radial direction,
particularly near the disc centre.

\begin{figure}
\includegraphics[width=\hsize,angle=0]{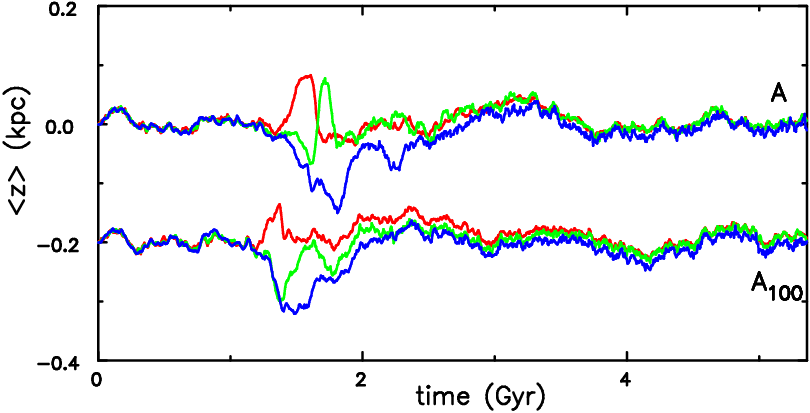}
\caption{As for Fig.~ref{fig.zbend} but for models A (reproduced from
  Fig.~\ref{fig.zbend}) and \Ab.  Values for model \Ab\ have been
  shifted by $-200$~pc for clarity.}
\label{fig.zbend_100}
\end{figure}

The first consequence of this change is that the bar instability is a
little more vigorous.  The bar in model \Ab\ formed by $t \sim
0.5\;$Gyr and buckled around $t \sim 1.4\;$Gyr, as shown in
Figure~\ref{fig.zbend_100}, a little earlier than in model A.  After
buckling, the evolution of the bar in model \Ab\ resembled that in A,
shown in Fig.~\ref{fig.4625_bar}, in length, amplitude, pattern speed,
and thickness (despite having started from a thinner disc).

The bar in model \Cb\ formed at the same time as that in model \Ab, as
expected, but its subsequent evolution was not as extreme as that in
model C.  It neither became as strong, nor as long, nor did it slow
down or puff up to the same extent as shown in
Fig.~\ref{fig.4645_bar}.  In this case the halo gained $\sim 22$\% of
the initial disc angular momentum, which is still a substantial
fraction, but considerably less than in model C over the same
interval.

\begin{figure*}
\includegraphics[width=.667\hsize,angle=0]{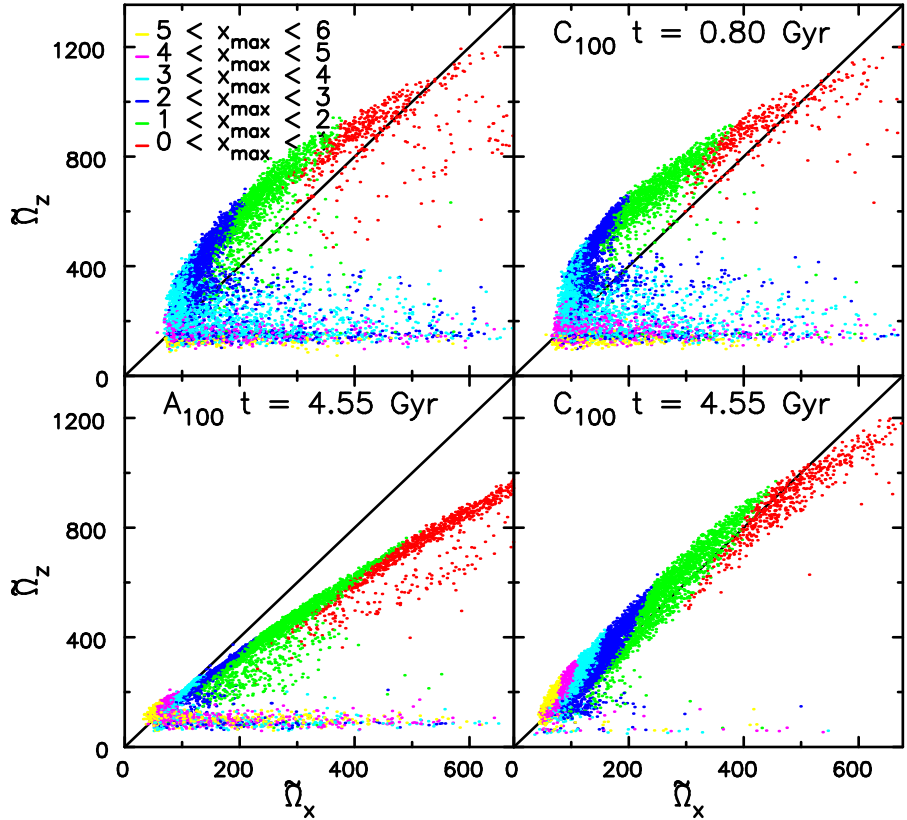}
\caption{As for Fig.~\ref{fig.frqplt} but for models \Ab\ (left) and
  \Cb\ (right).  The frequency ranges have been increased and the top
  row is for $t=0.8\;$Gyr.}
\label{fig.frqplt_100}
\end{figure*}

\subsection{Buckling instability}
Figure~\ref{fig.frqplt_100} presents the distributions in frequency
space of the particles in models \Ab\ (left) and \Cb\ (right), at
$t=0.8\;$Gyr (top) and $t=4.55\;$Gyr (bottom).  Notice that the
frequency ranges have been increased in comparison with those of
Fig.~\ref{fig.frqplt}.  Most points from particles within the bar
($x_{\rm max} \leq 3\;$kpc) in both both models now lie well above the
2:1 resonance line (marked by the diagonal), yet the bar in model
\Ab\ buckled in the same manner as in model A.  As argued in
\S\ref{sec.buckle} above, and originally by \citet{MS94}, the
collective instability requires that most particles have $\Omega_z >
2\Omega$, and this condition is still true in model \Ab\ even though
the particle distribution in model \Ab\ is clearly farther from the
resonance line in \Ab\ than it is in A.  Thus the buckling instability
does not require bar particles to be close to the 2:1 vertical
resonance as was also illustrated for a single orbit in the upper
right panel of Fig.~\ref{fig.orbit1286}.

It is also noteworthy that the distribution of points in the lower
left panel of Figure~\ref{fig.frqplt_100} (model \Ab) again is
strongly concentrated in a line of similar slope and intercept as that
in the lower left panel of Figure~\ref{fig.frqplt} (model A) although
now with a wider frequency range.

\subsection{Resonance passage {\it vs} resonance capture}
While it is true that the 2:1 vertical resonance plays a pivotal role
in the thickening of the bars in both models B and C, the mechanisms
are quite distinct.  \citet{Qu14} noted this distinction, but argued
that the trapping mechanism proposed by \citet{Qu02} was unlikely to
be of importance because ``the resonance width was too narrow''.  We
presented compelling evidence in \S\ref{sec.Q14test} that particles in
model B were heated vertically as the resonance swept past them
\citep{Qu14}.

The behaviour in model C is quite different.
Figure~\ref{fig.hifracts}(C) shows the time evolution of the fraction of
vertically heated orbits in model C.  The heated fractions are
generally lower than in model B, consistent with the differing density
profiles (Fig.~\ref{fig.rhozprof}).  But the more interesting contrast
with Fig.~\ref{fig.hifracts}(B), is that the bar particles in model C
were heated almost continuously at all radii.  Thus there is no
evidence that the resonance was sweeping outwards along the bar, but
instead vertical heating by capture into resonance was happening
steadily at all $x_{\rm max}$.  Note that vertical heating began for
$x_{\rm max}> 3$ (cyan) and $>4\;$kpc (magenta) at later times because
the bar grew in length over time (Fig.~\ref{fig.4645_bar}(b)).
Particles were heated vertically as they became trapped, and the
fraction of trapped orbits increased continually to the end of the
simulation.  Thus we have found a case in which the 2:1 resonance
extends over a large part of the bar that seems consistent with the
original suggestion in \citet{Qu02}.

The mid-plane density in the bar of model C was high enough to hold
the particles in resonance, once captured.  The lower density in model
B, prevented particles in the inner bar from remaining in resonance.

As noted above, the evolution of the bar in model \Cb\ differed
significantly from that in C, yet the distribution of particles in
frequency space in the bottom right panel of
Figure~\ref{fig.frqplt_100} again straddles the diagonal to about the
same extent as it did in the lower right panel of
Figure~\ref{fig.frqplt}, albeit over a slightly more extensive
frequency range in this new case.  Visual examination of the orbits of
particles in model \Cb\ again revealed that an increasing fraction
particles within the bar held the ratio $\Omega_z \simeq 2\Omega_x$,
after it was reached, also suggesting that they had become trapped
into the 2:1 vertical resonance.

\begin{figure}
\includegraphics[width=\hsize,angle=0]{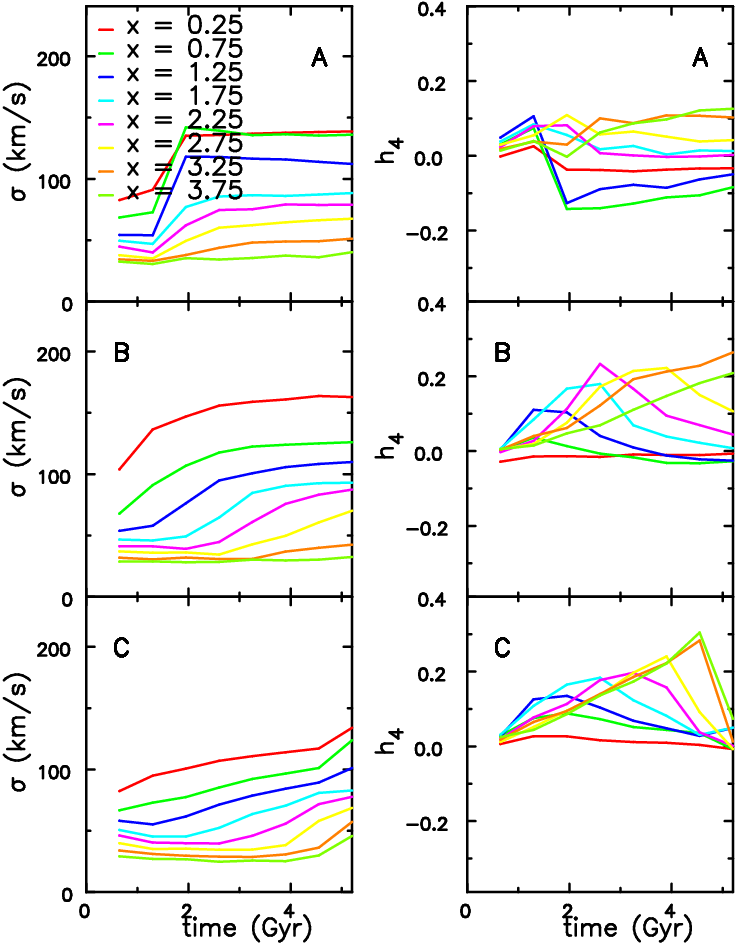}
\caption{The time evolution of the vertical velocity distributions of
  particles having $|z|<200\;$pc in the bars of model A (top) that did
  buckle, model B (middle) that did not buckle, and model C (bottom)
  in which buckling was suppressed.  The left panels show the dispersion
  $\sigma$ and the right the coefficient $h_4$.  The colours code the
  measurements in bins of different $x$, with the indicated mean
  values in kpc, and particles lying at distances from the bar major
  axis $y>1.6\;$kpc have also been excluded.}
\label{fig.GHevol}
\end{figure}

\section{Comparison with observations}
The bars in models A and B puffed up by different mechanisms: a
secular resonant trapping process for lifting orbits in model B and a
dynamical buckling instability in model A, although secular thickening
continued after the buckling instability had run its course.  The
final bars in these two cases are quite similar, as summarized in
Figures \ref{fig.4625_bar} and \ref{fig.4642_bar}.  The bar in model B
is slightly weaker (panels a), shorter (panels b) and not quite so
thick (panels d) as that in model A, but their pattern speeds (panels
c) and thickness profiles (panels d) were remarkably similar.  Also,
the overall thickening process occurs over similar periods, even
though the rapid surge associated with the saturation of the buckling
instability in model A, differs from the more gradual thickening in
model B.  A more significantly different bar developed in model C,
that also thickened by a third mechanism: resonant trapping.

With the objective to find fossil evidence in the velocity
distribution that could allow a bar that buckled to be distinguished
from one that did not, we have computed the coefficients $h_3$ and
$h_4$ of a Gauss-Hermite expansion \citep{Ge93, vF93} of the vertical
velocity distribution.  A non-zero value of $h_3$ would reflect a skew
velocity distribution, a positive value of $h_4$ indicates a velocity
distribution that is more peaked than a Gaussian and also has heavier
tails, while the distribution is more flat topped with weaker tails
when $h_4<0$.

We binned bar particles in slices of $|z|$-height, and in
$|x|$-distance along the bar, and determined these coefficients from
the particle velocities in each bin as described in the Appendix.

The results of this analysis for particles with $|z|<200\;$pc are
presented as functions of time in Figure~\ref{fig.GHevol}.  The
gradual rise of $\sigma$ in models B and C, compared with the sudden
surge for $x \la 2\;$kpc between $1.3 \leq t \leq 1.95\;$Gyr in model
A, reflects the thickening histories already presented in
Figs.~\ref{fig.4625_bar}, \ref{fig.4642_bar}, and \ref{fig.4645_bar}.
We do not show the evolution of $h_3$, because it remained near zero
in all three models.  However, there are clear differences in $h_4$:
on the one hand, the $h_4$ values in models B and C (middle and bottom
rows) tend to be positive, and one can observe the peak in $h_4$
gradually shifting to larger distance from the centre, with values
declining at each radius as the peak moves farther out.  On the other
hand, the buckling event in model A (top row) causes $h_4$ to become
strongly negative for $0.5 \leq |x| \leq 2\;$kpc (green and blue
lines), as was also reported by \citet{De05}, although the values in
these bins subsequently trend back towards zero gradually over time.
Farther out in the bar of model A, the behaviour more closely
resembles that in model B, reflecting the continued puffing up of the
bar, after the buckling event, by the resonance trapping mechanism at
later times.

These systematic differences between the $h_4$ values can be traced to
larger height above and below the mid-plane, although trends weaken
and the differences merge into the noise for $|z| \ga 500\;$pc.

These differences must be consequences of differences in the vertical
density profile of the bar.  When $h_4>0$, the velocity distribution
is more peaked than a Gaussian, and also has a heavy tail of high
velocity stars, which seems likely to reflect a vertical density
profile having a ``core-halo'' structure.  Conversely, the density
profile will be more truncated with height when $h_4<0$.
Figure~\ref{fig.rhozprof} presents the vertical density profiles in
the bars at the end of the evolution ($t=5.2\;$Gyr) in all three
models.  The density profile in model C near $|x|\sim 3.75\;$kpc has
heavy tails (cyan curve) where $h_4 > 0$, whereas it drops off more
steeply with height in models A and B near $|x|\sim 1.75\;$kpc when
$h_4 < 0$ (green curves), as expected from this argument.

We have verified that this same difference reflects the thickening
mechanisms in other models.  In particular, we have found strongly
negative values of $h_4$ near where the buckling instability produced
the peanut shape in another case, not presented here.  Even though the
$h_4$ values had all relaxed to near zero by the end of the evolution
in that model, the behaviour still contrasts with the large positive
values at some radii in the bars that puff up more gradually.

\subsection{Comparison with Milky Way}
We suggest that the sign and radial variation of the $h_4$ coefficient
of the vertical velocity distribution could be used as fossil
indicator of a bar that has buckled in its history.  In particular,
the values of $h_4$ for the vertical velocity distribution of stars in
the Milky Way could be used to test whether the bar in our Galaxy has
buckled in the past.  However, we defer this test to a second paper,
since extracting this information from the proper motions of the
red-clump giants of the Virac-Gaia survey \citep{Sm18, Cl19} requires
a detailed discussion.

\begin{figure}
\includegraphics[width=\hsize,angle=0]{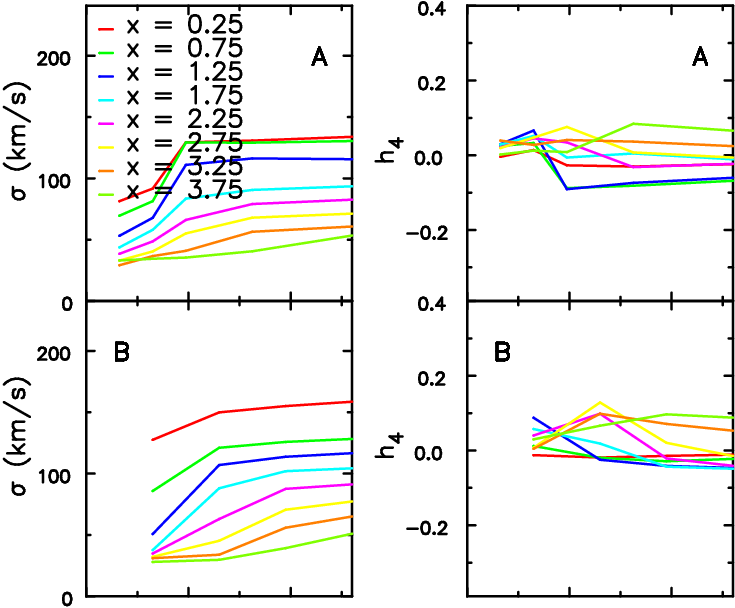}
\caption{The time evolution of the $z$-projected vertical velocity
  distributions of particles near the major axis of the bars of model
  A (above) that did buckle and model B (below) that did not buckle.
  The dispersion $\sigma$ is shown in the left panel and the
  coeffificent $h_4$ on the right.  The colours code measurements in
  bins of different $x$ distances along the bar, with the indicated
  mean values in kpc, and particles lying at distances from the bar
  major axis $y>0.4\;$kpc are eliminated.}
\label{fig.GHevolp}
\end{figure}

\subsection{Comparison with external galaxies}
The data presented in Fig~\ref{fig.GHevol} were for comparison with
the Milky Way, but we can also make predictions for projected velocity
distributions such as might be observed in external galaxies.
\citet{Me08} took long-slit spectra of two face-on barred galaxies,
and compute a Gauss-Hermite fit to the absorption line profiles.  We
therefore show in Fig~\ref{fig.GHevolp} the $z$-projected velocity
distribution of models A and B using a pseudo-slit of half-width
400~pc; the different colours are for different $x$-distances along
the bar.  Since the particle density is highest near the mid-plane, it
is no surprise that signal in the $h_4$ coefficients is only slightly
weaker than those from the mid-plane only, shown in
Fig~\ref{fig.GHevol}.

For NGC~98, \citet{Me08} found negative values of $h_4$ out to
5\arcsec\ from the centre along the bar major axis and concluded, on
the basis of comparison with simulations by \citet{De05}, that the bar
had buckled.  Our findings would support their conclusion.  However,
they reported positive values of $h_4$ in the case of NGC~600, for
which they offered no explanation.  Our models would suggest that if
the bar in that galaxy has thickened \citep[][suggest it may not
  have]{ED13}, it might have resulted from resonance heating.

\section{Conclusions}
\label{sec.concl}
We have presented simulations of bar formation and evolution in
idealized models of a disc embedded in a spherical halo.  In most
models, the evolution was computed self-consistently, but we have also
presented cases in which the potential was constrained throughout to
remain reflection symmetric about the mid-plane in order to suppress
the buckling instability.  The bars in all our models thickened, but
by three different mechanisms.

A buckling instability in a two-component disc-halo model caused some
of the thickening when the evolution was unrestricted.  We found that
adding a nuclear star cluster in the centre, having just 2.5\% of the
disc mass, seemed sufficient to suppress the buckling instability.  We
have presented substantial evidence that the bar in this case puffed
up because the $z$-excursions of orbits were increased as a 2:1
vertical resonance swept out along the bar, as was elegantly described
by \citet{Qu14}.  Particles were heated as they became resonant, and
the increased amplitude of their vertical motion persisted after they
left the resonance.

We have also uncovered a third thickening mechanism: gradual trapping
into the 2:1 vertical resonance.  In this mechanism, particles all
along the bar were heated as they become captured into the resonance,
where they remained until the end of the simulation.  \citet{Qu02} had
proposed this as a thickening mechanism, but no example has previously
been claimed, to our knowledge.  Note that we have observed this
mechanism only in our vertically symmetrized models, but it may
possibly also occur in unrestricted models having different initial
conditions.

We have reported only a few simulations, in which we have identified
the three different thickening mechanisms, but we have run many more
and have not noticed a fourth thickening mechanism among them.
However, we cannot exclude the possibility that other bars might
thicken by as yet undiscovered means.

Since we are unable to observe the evolution of any individual galaxy,
the mechanism by which the bar thickened can be deduced only from
fossil evidence \citep[unless the bar is observed directly in the
  short buckling phase, as was claimed by][]{ED16}.  We have found
clear evidence that the Gauss-Hermite coefficient $h_4$ of the
vertical velocity distribution provides such fossil evidence of a
bar's dynamical history: it is negative in the inner regions of a
buckled bar, as previously found by \citet{De05}, but here we show
that it is positive when the bar has been puffed up by either of the
two more gradual mechanisms.  We caution that we have found this
distinction in rather few bar models, and so cannot say that it is
completely general.

\citet{Me08} present evidence for negative $h_4$ values from the
vertical velocity distribution of the face-on bar of NGC~98, which
they argued was evidence that the bar had buckled.  But they also
found positve $h_4$ values in NGC~600; if the bar in this galaxy has
puffed up, our work suggests that the mechanism proposed by
\citet{Qu14} could be responsible.

We are currently applying this diagnostic to the vertical velocity
distribution of stars in the Milky Way bulge, which will the subject
of a subsequent paper.

\section*{Acknowledgements}
We thank Victor Debattista and Jonny Clarke for helpful comments and
an anonymous referee for a prompt report that encouraged us to provide
more evidence to support our case.  This work was begun during visits
by JAS to MPE Garching in 2016 and 2017 that were supported by the DFG
cluster of excellence ``Origin and Structure of the Universe.''  JAS
also acknowledges the continuing hospitality of Steward Observatory.


\def\erf{\hbox{erf}\,}
\begin{onecolumn}
\section*{Appendix}
\appendix
\citet{vF93} and \citet{Ge93} noted that a non-Gaussian line profile
may be described as a Gauss-Hermite series
\begin{equation}
{\cal L}(v) = \left[ {A\alpha(w) \over \sigma} \right] \sum_k h_kH_k(w), \qquad\hbox{where}\qquad w= {v-V \over \sigma}.
\label{eq.GHexp}
\end{equation}
Here, the standard Gaussian function $\alpha(w) =
\exp(-w^2/2)/\sqrt{2\pi}$, $V$ and $\sigma$ are the mean and
dispersion of the best fitting Gaussian, and $A$ is its
normalization.  The functions $H_k(w)$ are Hermite polynomials, and
the coefficients $h_k$ are free parameters.  When $\alpha$ is defined
in this standard way, the first seven polynomials are
\begin{eqnarray}
H_0(w) = 1, \qquad
H_1(w) = \sqrt{2}w, & \qquad &
H_2(w) = (2w^2 - 1 )/\sqrt{2}, \nonumber \\
H_3(w) = (2w^3 - 3w)/\sqrt{3}, & \qquad &
H_4(w) = (4w^4 - 12w^2 + 3)/\sqrt{24},  \\
H_5(w) = (4w^5 - 20w^3 + 15w)/\sqrt{60},
& \qquad\hbox{and}\qquad &
H_6(w) = ( 8w^6 - 60w^4 + 90w^2 - 15 ) / \sqrt{720}. \nonumber
\end{eqnarray}
These expressions are normalized such that $\int_{-\infty}^\infty
\alpha(w) H_k(w) \; \alpha(w) H_m(w) \; dw =
\delta_{km}(4\pi)^{-1/2}$, where $\delta_{km}$ is the Kronecker delta,
in agreement with eq.~(6) of \citet{vF93}.  Since these functions are
orthogonal, the coefficients $h_k$ in eq.~(\ref{eq.GHexp}) are
independent of each other.

Following \citet{vF93}, we hold $(h_0,h_1,h_2)$ fixed to to $(1,0,0)$,
while allowing $(A,V,\sigma)$ to be free parameters of the fit to the
generalized profile; \citet{MB94} prove that this simplification does
not alter the fit.  Thus, a fit to the line profile for $k \leq 4$
has the functional form
\begin{equation}
{\cal L}(v) = \left[ {A\alpha(w) \over \sigma} \right] \left\{ 1 + h_3H_3(w) + h_4H_4(w) \right\},
\label{eq.GHline}
\end{equation}
with five free parameters: $(A, V, \sigma, h_3, h_4)$.

It should be noted that the function (\ref{eq.GHline}) becomes
negative over some range(s) of $v$ for any non-zero value of $h_3$ and
all but small positive values of $h_4$.  When both coefficients have
small absolute values, the range(s) where the function is negative is
(are) out in the wings of the profile where the Gaussian factor causes
negative values of the function to have little significance.  However,
when fitting this function to the velocity distribution of particles,
the values of $h_3$ and $h_4$ will be constrained by the fact that the
function to be fitted is nowhere negative.

We could construct an ``observed'' line profile, ${\cal L}_0(v)$ from
the velocity distribution of a finite number of particles, and then
try to fit the expression (\ref{eq.GHline}) to it by a least squares
minimization for ${\cal S}_{\rm s} = \sum_j \left[{\cal L}_o(v_j) -
  {\cal L}(v_j)\right]^2$, at some set of velocities $\{v_j\}$.  This
would be undesirable, however, because we would need to define a
smooth function ${\cal L}_o(v)$ through a kernel estimate, for
example, which would immediately introduce a bias.

Fitting the cumulative velocity distribution avoids this particular
bias. (The values of $h_3$ and $h_4$ will continue to be constrained
by the always increasing cumulative distribution).  The integral of
the line profile (\ref{eq.GHline}) to velocity $v$, or to $w$, is
\begin{eqnarray}
{\cal N}(v) & = & \int_{-\infty}^v{\cal L}(v^\prime) \; dv^\prime = {A \over \sigma} \int_{-\infty}^w \alpha( w^\prime ) 
\left\{ 1 + h_3H_3(w^\prime) + h_4H_4(w^\prime) \right\} dw^\prime \nonumber \\
& = & {A \over \sigma} \left\{ Q_0(w) + {h_3 \over \sqrt 3}\left[2Q_3(w) - 3Q_1(w)\right] + {h_4 \over \sqrt{24}}\left[4Q_4(w) -12Q_2(w) +3 Q_0(w) \right] \right\}.
\end{eqnarray}
Since $\erf(x) \equiv (2 / \pi^{1/2}) \int_0^x e^{-t^2} dt$, we find
\begin{equation}
Q_l(w) = \int_{-\infty}^w w^{\prime l} \alpha(w^\prime)\; dw^\prime = \begin{cases}
 \displaystyle{1\over2}\left[1 + \erf\left({w\over\surd2}\right)\right] & l=0 \\
 -\alpha(w) \phantom{\displaystyle{1\over2}} & l=1  \\
 \displaystyle{1 \over 2} \left[1 + \erf\left({w\over\surd2}\right)\right] - w\alpha(w) & l=2 \\
 -\alpha(w)\left(w^2+2\right) \phantom{\displaystyle{1\over2}} & l=3 \\
 \displaystyle{3\over 2}\left[1 + \erf\left({w\over\surd2}\right)\right] - w\alpha(w) \left(w^2 + 3\right). & l=4 \\
\end{cases}
\end{equation}
The expression for $Q_0$ is standard, the integrals for $l=1$ and
$l=3$ are easy, while those for $l=2$ and $l=4$ must be integrated by
parts with the help of formula 5.41 from \citet{GR80}.

Now assume we have a set of $n$ discrete velocities, $\{v_i\}$.  The
cumulative distribution of the input data to be fitted is therefore
\begin{equation}
{\cal N}_o(v_i) = {1 \over n} \sum_{j=1}^n H(v_i - v_j),
\label{eq.inpdata}
\end{equation}
where the Heaviside function $H(v_i-v_j) = 1$ if $v_i\geq v_j$ and
$H(v_i-v_j) = 0$ otherwise.  The best fit line profile is that for
which the five parameters $(A, V, \sigma, h_3, h_4)$ have the values
that minimize
\begin{equation}
{\cal S}_{\rm u} = \sum_{i=1}^n \left[{\cal N}_o(v_i) - {\cal N}(v_i)\right]^2.
\end{equation}
Clearly, $A/\sigma$ must be close to unity, because of the
normalization of ${\cal N}_o$ in eq.~(\ref{eq.inpdata}).  We use the
fitting tool, {\tt sumsl}, \citep[from the software collection made
  public by][]{Bu17} that finds the set of five parameters $\{p_j\}$
that minimize ${\cal S}_{\rm u}$.  Note that this tool requires a
routine to supply a vector of gradients $\{\partial{\cal S}_u/\partial
p_j\}$ at any point on the hyper-surface.

As a measure of the goodness of the fit, we use a Kolmogorov-Smirnov
test to estimate the probability that the data $\{v_i\}$ were drawn
from the fitted distribution ${\cal N}(v)$.

It might be objected that the integrated Hermite polynomials
$\int_{-\infty}^w \alpha(w^\prime)H_k(w^\prime)\; dw^\prime$ are not
orthogonal, and the coefficients $h_k$ will therefore no longer be
independent.  Although this is true, we find the best fit values of
$h_3$ and $h_4$ derived by minimizing ${\cal S}_{\rm u}$ using the
unsmoothed cumulative distribution agree quite well with those
obtained from minimizing ${\cal S}_{\rm s}$ to the smoothed data.  The
main source of difference is that smoothing reduces the magnitude of
$h_3$ and $h_4$, as is to be expected.  In all cases that we have
examined, the KS probability that the values were drawn from the
fitted distribution were higher for the cumulative fit than for the
fit to the smoothed ``line profile.''

\end{onecolumn}

\bsp	
\label{lastpage}

\begin{thebibliography}{99}
\def\skip#1{ \etal\ }
\def\PhD{PhD thesis.}
\def\rmp{Rev. Mod. Phys.}
\def\rpp{Rep. Prog. Phys.}

\bibitem[\protect\citeauthoryear{Abbott \etal}{2017}]{Ab17}
Abbott, C. G., Valluri, M., Shen, J. \& Debattista, V. P. 2017, \mnras, {bf 470}, 1526

\bibitem[\protect\citeauthoryear{Aguerri \etal}{2015}]{Ag15}
Aguerri, J. A. L., M\'endez-Abreu, J., Falc\'on-Barroso, J.,\skip{ Amorin, A., Barrera-Ballesteros, J., Cid Fernandes, R., Garc\'\i a-Benito, R., Garc\'\i a-Lorenzo, B., Gonz\'alez Delgado, R. M., Husemann, B., Kalinova, V., Lyubenova, M., Marino, R. A., M\'arquez, I., Mast, D., P\'erez, E., S\'anchez, S. F., van de Ven, G., Walcher, C. J., Backsmann, N. Cortijo-Ferrero, C., Bland-Hawthorn, J., del Olmo, A., Iglesias-P\'aramo, J., P\'erez, I., S\'anchez-Bl\'azquez, P., Wisotzki, L. \& Ziegler, B.} 2015, \aap, {\bf 576}, A102

\bibitem[\protect\citeauthoryear{Binney, Gerhard \& Spergel}{1997}]{Bi97}
Binney, J., Gerhard, O. \& Spergel, D. 1997, \mnras, {\bf 288}, 365

\bibitem[\protect\citeauthoryear{Binney \& Tremaine}{2008}]{BT08}
Binney, J. \& Tremaine, S. 2008, {\it Galactic Dynamics\/} 2nd Ed.\ (Princeton: Princeton University Press)

\bibitem[\protect\citeauthoryear{Bland-Hawthorn \& Gerhard}{2016}]{BHG16}
Bland-Hawthorn, J. \& Gerhard, O. 2016, \araa, {\bf 54}, 529

\bibitem[\protect\citeauthoryear{Bureau \& Athanassoula}{2005}]{BA05}
Bureau, M. \& Athanassoula, E. 2005, \apj, {\bf 626}, 159

\bibitem[\protect\citeauthoryear{Burkardt}{2017}]{Bu17}
Burkardt, J. 2017, \\ Website: {\tt https://people.sc.fsu.edu/$\sim$jburkardt/}

\bibitem[\protect\citeauthoryear{Clarke \etal}{2019}]{Cl19}
Clarke, J. P., Wegg, C., Gerhard, O., Smith, L., Lucas, P. \& Wylie, S. 2019, \mnras, {\bf 489}, 3519

\bibitem[\protect\citeauthoryear{Colin \& Athanassoula}{1989}]{CA89}
Colin, J. \& Athanassoula, E. 1989, \aap, {\bf 214}, 99

\bibitem[\protect\citeauthoryear{Collier}{2020}]{Co20}
Collier, A. 2020, \mnras, {\bf 492}, 2241

\bibitem[\protect\citeauthoryear{Combes \etal}{1990}]{Co90}
Combes, F., Debbasch, F., Friedli, D. \& Pfenniger, D. 1990, \aap, {\bf 233}, 82

\bibitem[\protect\citeauthoryear{Combes \& Sanders}{1981}]{CS81}
Combes, F. \& Sanders, R. H. 1981, \aap, {\bf 96}, 164

\bibitem[\protect\citeauthoryear{Dame, Hartmann \& Thaddeus}{2001}]{DHT01}
Dame, T., Hartmann, D. \& Thaddeus P. 2001. \apj, {\bf 547}, 792

\bibitem[\protect\citeauthoryear{Debattista \etal}{2005}]{De05}
Debattista, V. P., Carollo, C. M., Mayer, L. \& Moore, B. 2005, \apj, {\bf 628}. 678

\bibitem[\protect\citeauthoryear{Debattista \& Sellwood}{2000}]{DS00}
Debattista, V. P. \& Sellwood, J. A. 2000, \apj, {\bf 543}, 704

\bibitem[\protect\citeauthoryear{de Vaucouleurs \& Freeman}{1972}]{dVF72}
de Vaucouleurs, G. \& Freeman, K. C. 1972, {\it Vistas Astron.}, {\bf 14}, 163

\bibitem[\protect\citeauthoryear{D\'\i az-Garc\'\i a \etal}{2016}]{DG16}
D\'\i az-Garc\'\i a, S., Salo, H., Laurikainen, E. \& Herrera-Endoqui, M. 2016, \aap, {\bf 587}, A160

\bibitem[\protect\citeauthoryear{Erwin}{2018}]{Er18}
Erwin, P. 2018, \mnras, {\bf 474}, 5372

\bibitem[\protect\citeauthoryear{Erwin \& Debattista}{2013}]{ED13}
Erwin, P. \& Debattista, V. P. 2013, \mnras, {\bf 431}, 3060

\bibitem[\protect\citeauthoryear{Erwin \& Debattista}{2016}]{ED16}
Erwin, P. \& Debattista, V. P. 2016, \apjl, {\bf 825}, L30

\bibitem[\protect\citeauthoryear{Erwin \& Debattista}{2017}]{ED17}
Erwin, P. \& Debattista, V. P. 2017, \mnras, {\bf 468}, 2058

\bibitem[\protect\citeauthoryear{Friedli \& Pfenniger}{1990}]{FP90}
Friedli, D. \& Pfenniger, D. 1990, in ``Bulges of Galaxies'', eds. B. J. Jarvis \& D. M. Terndrup (Garching: ESO workshop {\bf 35}) p.~265

\bibitem[\protect\citeauthoryear{Fux}{1999}]{Fu99}
Fux, R. 1999, \aap, {\bf 345}, 787

\bibitem[\protect\citeauthoryear{Gajda \etal}{2016}]{GLA16}
Gajda, G., \L okas, E. L. \& Athanassoula, E. 2016, \apj, {\bf 830}, 108

\bibitem[\protect\citeauthoryear{Gerhard}{1993}]{Ge93}
Gerhard, O. E. 1993, \mnras, {\bf 265}, 213

\bibitem[\protect\citeauthoryear{Gradshteyn \& Ryzhik}{1980}]{GR80}
Gradshteyn, I. S. \& Ryzhik, I. M. 1980, {\it Table of Integrals, Series, and Products\/} 4th Ed.\ (Orlando: Academic Press Inc.)

\bibitem[\protect\citeauthoryear{Hammersley \etal}{2000}]{Ha00}
Hammersley, P. L,, Garz\'on, F,, Mahoney, T.J., L\'opez-Corredoira, M. \& Torres, M.A.P. 2000, \mnras, {\bf 317}, L45

\bibitem[\protect\citeauthoryear{Hernquist}{1990}]{He90}
Hernquist, L. 1990, \apj, {\bf 356}, 359

\bibitem[\protect\citeauthoryear{Holmes \etal}{2015}]{Ho15}
Holmes, L., Spekkens, K., S\'anchez, S. F.,\skip{ Walcher, C. J., Garc\'\i a-Benito, R., Mast, D., Cortijo-Ferrero, C., Kalinova, V., Marino, R. A., Mendez-Abreu, J. \& Barrera-Ballesteros, J. K.} 2015, \mnras, {\bf 451}, 4397

\bibitem[\protect\citeauthoryear{Howard \etal}{2009}]{Ho09}
Howard, C. D., Rich, R. M., Clarkson, W.,\skip{ Mallery, R., Kormendy, J., De Propris, R., Robin, A. C., Fux, R., Reitzel, D. B., Zhao, HS., Kuijken, K. \& Koch, A.} 2009, \apjl, {\bf 702}, L153

\bibitem[\protect\citeauthoryear{Kormendy}{1983}]{Ko83}
Kormendy, J. 1983, \apj, {\bf 275}, 529

\bibitem[\protect\citeauthoryear{Kuijken \& Merrifield}{1995}]{KM95}
Kuijken, K. \& Merrifield, M. R. 1995, \apjl, {\bf 443}, L13

\bibitem[\protect\citeauthoryear{Laskar}{1990}]{La90}
Laskar, J. 1990, Icar., {\bf 88}, 266

\bibitem[\protect\citeauthoryear{Li \etal}{2016}]{Li16}
Li, Z., Gerhard, O., Shen, J., Portail, M. \& Wegg, C. 2016, \apj, {\bf 824}, 13

\bibitem[\protect\citeauthoryear{\L okas}{2019}]{Lo19}
\L okas, E. L. 2019, \aap, {\bf 629}, A52

\bibitem[\protect\citeauthoryear{L\"utticke, Dettmar \& Pohlen}{2000}]{Lu00}
L\"utticke, R., Dettmar, R.-J. \& Pohlen, M. 2000, \aaps, {\bf 145}, 405

\bibitem[\protect\citeauthoryear{Magorrian \& Binney}{1994}]{MB94}
Magorrian, J. \& Binney, J. 1994, \mnras, {\bf 271}, 949

\bibitem[\protect\citeauthoryear{Martinez-Valpuesta \& Shlosman}{2004}]{MVS04}
Martinez-Valpuesta, I. \& Shlosman, I. 2004, \apjl, {\bf 613}, L29

\bibitem[\protect\citeauthoryear{Martinez-Valpuesta \etal}{2006}]{MV06}
Martinez-Valpuesta, I., Shlosman, I. \& Heller, C. 2006, \apj, {\bf 637}, 214

\bibitem[\protect\citeauthoryear{McWilliam \& Zoccali}{2010}]{MZ10}
McWilliam, A. \& Zoccali, M. 2010, \apj, {\bf 724}, 1491

\bibitem[\protect\citeauthoryear{M\'endez-Abreu \etal}{2008}]{Me08}
M\'endez-Abreu, J., Corsini, E. M., Debattista, Victor P., De Rijcke, S., Aguerri, J. A. L. \& Pizzella, A. 2008, \apjl, {\bf 679}, L73

\bibitem[\protect\citeauthoryear{Merritt \& Sellwood}{1994}]{MS94}
Merritt, D. \& Sellwood, J. A. 1994, \apj, {\bf 425}, 551

\bibitem[\protect\citeauthoryear{Ness \& Lang}{2016}]{NL16}
Ness, M. \& Lang, D. 2016, \aj, {\bf 152}, 14

\bibitem[\protect\citeauthoryear{Petersen \etal}{2016}]{PWK16}
Petersen, M. S., Weinberg, M. D. \& Katz, N. 2016, \mnras, {\bf 463}, 1952

\bibitem[\protect\citeauthoryear{Petersen \etal}{2019a}]{PWK19a}
Petersen, M. S., Weinberg, M. D. \& Katz, N. 2019a, arXiv:1902.05081

\bibitem[\protect\citeauthoryear{Petersen \etal}{2019b}]{PWK19b}
Petersen, M. S., Weinberg, M. D. \& Katz, N. 2019b, \mnras, {\bf 490}, 3616

\bibitem[\protect\citeauthoryear{Portail \etal}{2017}]{Po17}
Portail, M., Gerhard, O., Wegg, C. \& Ness, M. 2017, \mnras, {\bf 465}, 1621

\bibitem[\protect\citeauthoryear{Portail \etal}{2015}]{Po15}
Portail, M., Wegg, C. \& Gerhard, O. 2015, \mnras, {\bf 450}, L66

\bibitem[\protect\citeauthoryear{Quillen}{2002}]{Qu02}
Quillen, A. C. 2002, \aj, {\bf 124}, 722

\bibitem[\protect\citeauthoryear{Quillen \etal}{2014}]{Qu14}
Quillen, A. C., Minchev, I., Sharma, S., Qin, Y-J. \& Di Matteo, P. 2014, \mnras, {\bf 437}, 1284

\bibitem[\protect\citeauthoryear{Raha \etal}{1991}]{Ra91}
Raha, N., Sellwood, J. A., James, R. A. \& Kahn, F. D. 1991, \nat, {\bf 352}, 411

\bibitem[\protect\citeauthoryear{Rangwala \etal}{2009}]{Ra09}
Rangwala, N., Williams, T. B. \& Stanek, K. Z. 2009, \apj, {\bf 691}, 1387

\bibitem[\protect\citeauthoryear{Sanders \etal}{2019}]{Sa19}
Sanders, J., Smith, L., Evans, N. W. \& Lucas, P. 2019, \mnras, {\bf 487}, 5188

\bibitem[\protect\citeauthoryear{Sellwood}{1989}]{Se89}
Sellwood, J. A. 1989, \mnras, {\bf 238}, 115

\bibitem[\protect\citeauthoryear{Sellwood}{2014}]{Se14}
Sellwood, J. A. 2014, arXiv:1406.6606 (on-line manual: \hfil\break {\tt http://www.physics.rutgers.edu/$\sim$sellwood/manual.pdf})

\bibitem[\protect\citeauthoryear{Sellwood \& McGaugh}{2005}]{SM05}
Sellwood, J. A. \& McGaugh, S. S. 2005, \apj, {\bf 634}, 70

\bibitem[\protect\citeauthoryear{Sellwood \& Merritt}{1994}]{SM94}
Sellwood, J. A. \& Merritt, D. 1994, \apj, {\bf 425}, 530

\bibitem[\protect\citeauthoryear{Sellwood \& Wilkinson}{1993}]{SW93}
Sellwood, J. A. \& Wilkinson, A. 1993, \rpp, {\bf 56}, 173

\bibitem[\protect\citeauthoryear{Shaw}{1987}]{Sh87}
Shaw, M. A. 1987, \mnras, {\bf 229}, 691

\bibitem[\protect\citeauthoryear{Shen \etal}{2010}]{Sh10}
Shen, J., Rich, R. M., Kormendy, J.,\skip{ Howard, C. D., De Propris, R. \& Kunder, A.} 2010, \apjl, {\bf 720}, L72

\bibitem[\protect\citeauthoryear{Smith \etal}{2018}]{Sm18}
Smith, L. C., Lucas, P. W., Kurtev, R.,\skip{ Smart, R., Minniti, D., Borissova, J., Jones, H. R. A., Zhang, Z. H., Marocco, F., Contreras Pe\~na, C., Gromadzki, M., Kuhn, M. A., Drew, J. E., Pinfield, D. J. \& Bedin, L. R.} 2018, \mnras, {\bf 474}, 1826

\bibitem[\protect\citeauthoryear{Sparke \& Sellwood}{1987}]{SS87}
Sparke, L. S. \& Sellwood, J. A. 1987, \mnras, {\bf 225}, 653

\bibitem[\protect\citeauthoryear{Toomre}{1981}]{To81}
Toomre, A. 1981, In ''The Structure and Evolution of Normal Galaxies'', eds.~S. M. Fall \& D. Lynden-Bell (Cambridge, Cambridge Univ. Press) p.~111

\bibitem[\protect\citeauthoryear{Valluri \etal}{2016}]{Va16}
Valluri, M., Shen, J., Abbott, C. \& Debattista, V. P. 2016, \apj, {\bf 818}, 141

\bibitem[\protect\citeauthoryear{van der Marel \& Franx}{1993}]{vF93}
van der Marel, R. \& Franx, M. 1993, \apj, {\bf 407}, 525

\bibitem[\protect\citeauthoryear{Weiland \etal}{1994}]{We94}
Weiland, J. L., Arendt, R. G., Berriman, G. B.,\skip{ Dwek, E., Freudenreich, H. T., Hauser, M. G., Kelsall, T., Lisse, C. M., Mitra, M., Moseley, S. H., Odegard, N. P., Silverberg, R. F., Sodroski, T. J., Spiesman, W. J. \& Stemwedel, S. W.} 1994, \apjl, {\bf 425}, L81

\bibitem[\protect\citeauthoryear{Wegg \& Gerhard}{2013}]{WG13}
Wegg, C. \& Gerhard, O. 2013, \mnras, {\bf 435}, 1874

\bibitem[\protect\citeauthoryear{Wegg \etal}{2015}]{We15}
Wegg, C., Gerhard, O. \& Portail, M. 2015, \mnras, {\bf 450}, 4050

\bibitem[\protect\citeauthoryear{Weiner \etal}{2001}]{We01}
Weiner, B. J., Williams, T. B., van Gorkom, J. H. \& Sellwood, J. A. 2001, \apj, {\bf 546}, 916

\end{thebibliography}
\end{document}